\documentclass[10pt, letterpaper]{IEEEtran}
\IEEEoverridecommandlockouts
\usepackage{cite}
\usepackage{amsmath,amssymb,amsthm}
\usepackage{bbm}
\usepackage{algpseudocode}
\usepackage{graphicx}
\usepackage{textcomp}
\usepackage{mathtools}
\usepackage{xcolor}
\usepackage{algorithm}
\usepackage{bm}
\usepackage{comment}

\algnewcommand\algorithmicswitch{\textbf{switch}}
\algnewcommand\algorithmiccase{\textbf{case}}
\algdef{SE}[SWITCH]{Switch}{EndSwitch}[1]{\algorithmicswitch\ #1\ \algorithmicdo}{\algorithmicend\ \algorithmicswitch}%
\algdef{SE}[CASE]{Case}{EndCase}[1]{\algorithmiccase\ #1}{\algorithmicend\ \algorithmiccase}%
\algtext*{EndSwitch}%
\algtext*{EndCase}%

\def\BibTeX{{\rm B\kern-.05em{\sc i\kern-.025em b}\kern-.08em
    T\kern-.1667em\lower.7ex\hbox{E}\kern-.125emX}}
\usepackage{url}

\usepackage{tikz}
\usetikzlibrary{shapes, arrows, positioning,plotmarks,patterns}

\usepackage{subfig}

\usepackage{pgfplots}
\pgfplotsset{
compat=1.3,
legend style={font=\footnotesize, fill opacity=0.7,  draw opacity=1, text opacity=1, draw=white!15!black, legend cell align=left, align=left}, 
width=6cm, 
height=6cm,
yminorticks=false,
xminorticks=false,
title style={font=\small},
tick style={color=black},
tick label style={font=\scriptsize},
xlabel style={font=\footnotesize},
ylabel style={font=\footnotesize},
grid style={line width=.1pt, draw=gray!50},
major grid style={line width=.1pt,draw=gray!50},
}
\usepgfplotslibrary{colorbrewer}
\usepgfplotslibrary{groupplots}
\usepgfplotslibrary{fillbetween}
\usetikzlibrary{patterns}
\pgfplotsset{every tick label/.append style={font=\footnotesize}}

\usepackage{glossaries}
\newacronym{zw}{ZW}{Zero-Wait}
\newacronym{cr}{CR}{Collision Resolution}
\newacronym{ce}{CE}{Collision Exit}
\newacronym{bt}{BT}{Belief Threshold}
\newacronym{aoii}{AoII}{Age of Incorrect Information}
\newacronym{aoi}{AoI}{Age of Information}
\newacronym{voi}{VoI}{Value of Information}
\newacronym{cdf}{CDF}{Cumulative Distribution Function}
\newacronym{pmf}{PMF}{Probability Mass Function}
\newacronym{delta}{DELTA}{Dynamic Epistemic Logic for Tracking Anomalies}
\newacronym{del}{DEL}{Dynamic Epistemic Logic}
\newacronym{tdd}{TDD}{Time Division Duplex}
\newacronym{nack}{NACK}{Negative ACK}
\newacronym{iot}{IoT}{Internet of Things}

\newacronym{rr}{RR}{Round-Robin}
\newacronym{maf}{MAF}{Maximum Age First}
\newacronym{lzw}{LZW}{Local Zero-Wait}
\newacronym{gzw}{GZW}{Global Zero-Wait}

\newcommand{\T}{^{\intercal}}     
\newcommand{\E}[1]{\mathbb{E}\left[ #1 \right]} 
\newcommand{\mc}[1]{\mathcal{#1}}   
\newcommand{\mb}[1]{\mathbf{#1}}    
\DeclareMathOperator*{\argmax}{arg\,max}    
\DeclareMathOperator*{\Geo}{Geo}    
\DeclareMathOperator*{\Int}{int}    
\DeclareMathOperator*{\Mod}{mod}    
\DeclareMathOperator*{\Bin}{Bin}    
\DeclareMathOperator*{\argmin}{arg\,min}    

\algtext*{EndWhile}
\algtext*{EndIf}
\algtext*{EndFor}

\def \fwidth{0.96\columnwidth}
\def \fheight {0.5\columnwidth}
\def \ffheight {0.48\columnwidth}
\def \ffwidth {0.48\columnwidth}

\definecolor{color3}{HTML}{2E15C8}
\definecolor{color2}{HTML}{BE37A4}
\definecolor{color1}{HTML}{F58275}
\definecolor{color0}{HTML}{FFD700}
\definecolor{barcolor0}{HTML}{00429D}
\definecolor{barcolor1}{HTML}{844D99}
\definecolor{barcolor2}{HTML}{C3608E}
\definecolor{barcolor3}{HTML}{EF8078}
\definecolor{barcolor4}{HTML}{FFB047}
\definecolor{darkslategray38}{RGB}{38,38,38}

\newtheorem{theorem}{Theorem}

\newtheorem{lemma}{Lemma}[theorem]

\makeatletter

\pgfdeclareshape{slash underlined}
{
  \inheritsavedanchors[from=rectangle] 
  \inheritanchorborder[from=rectangle]
  \inheritanchor[from=rectangle]{north}
  \inheritanchor[from=rectangle]{north west}
  \inheritanchor[from=rectangle]{north east}
  \inheritanchor[from=rectangle]{center}
  \inheritanchor[from=rectangle]{west}
  \inheritanchor[from=rectangle]{east}
  \inheritanchor[from=rectangle]{mid}
  \inheritanchor[from=rectangle]{mid west}
  \inheritanchor[from=rectangle]{mid east}
  \inheritanchor[from=rectangle]{base}
  \inheritanchor[from=rectangle]{base west}
  \inheritanchor[from=rectangle]{base east}
  \inheritanchor[from=rectangle]{south}
  \inheritanchor[from=rectangle]{south west}
  \inheritanchor[from=rectangle]{south east}
  \inheritanchorborder[from=rectangle]
  \foregroundpath{
    \southwest \pgf@xa=\pgf@x \pgf@ya=\pgf@y
    \northeast \pgf@xb=\pgf@x \pgf@yb=\pgf@y
    \pgf@xc=\pgf@xa
    \advance\pgf@xc by .5\pgf@xb
    \pgf@yc=\pgf@ya
    \advance\pgf@xc by -1.3pt
    \advance\pgf@yc by -1.8pt
    \pgfpathmoveto{\pgfqpoint{\pgf@xc}{\pgf@yc}}
    \advance\pgf@xc by  2.6pt
    \advance\pgf@yc by  3.6pt
    \pgfpathlineto{\pgfqpoint{\pgf@xc}{\pgf@yc}}
    \pgfpathmoveto{\pgfqpoint{\pgf@xa}{\pgf@ya}}
    \pgfpathlineto{\pgfqpoint{\pgf@xb}{\pgf@ya}}
 }
}
\tikzset{nomorepostaction/.code=\let\tikz@postactions\pgfutil@empty}

\title{Goal-Oriented Medium Access with \\ Distributed Belief Processing}
\author{Federico Chiariotti, \IEEEmembership{Senior Member, IEEE}, Andrea Munari, \IEEEmembership{Senior Member, IEEE},\\ Leonardo Badia, \IEEEmembership{Senior Member, IEEE}, and Petar Popovski, \IEEEmembership{Fellow, IEEE}\thanks{Federico Chiariotti (corresponding author, federico.chiariotti@unipd.it) and Leonardo Badia (leonardo.badia@unipd.it) are with the Department of Information Engineering, University of Padova, Padua, Italy. Andrea Munari (andrea.munari@dlr.de) is with the Institute of Communications and Navigation, German Aerospace Center (DLR), We{\ss}ling, Germany. Petar Popovski (petarp@es.aau.dk) is with the Department of Electronic Systems, Aalborg University, Aalborg \O{}st, Denmark. This work was supported in part by the Italian National Recovery and Resilience Plan (NRRP), as part of the RESTART partnership (PE0000001), under the European Union NextGenerationEU Project, by the Federal Ministry of Education and Research of Germany in the programme of ”Souverän. Digital. Vernetzt.” joint project 6G-RIC, project identification number: 16KISK022, and by the Villum Investigator Grant ``WATER'' from the Velux Foundations, Denmark. A preliminary version of this work has been presented at the IEEE INFOCOM 2025 conference~\cite{chiariotti2025delta}.}}

\begin{document}

\maketitle

\begin{abstract}
Goal-oriented communication entails the timely transmission of updates related to a specific goal defined by the application. In a distributed setup with multiple sensors, each individual sensor knows its own observation and can determine its freshness, as measured by \gls{aoii}. This local knowledge is suited for distributed medium access, where the transmission strategies have to deal with collisions.
We present \gls{delta}, a medium access protocol that limits collisions and minimizes \gls{aoii} in anomaly reporting over dense networks. Each sensor knows its own \gls{aoii}, while it can compute the belief about the \gls{aoii} for all other sensors, \emph{based on their \gls{aoi}}, which is inferred from the acknowledgments. This results in a goal-oriented approach based on dynamic epistemic logic emerging from public information. 
We analyze the resulting \gls{delta} protocol both from a theoretical standpoint and with Monte Carlo simulations, showing that it is significantly more efficient and robust than classical random access, while outperforming state-of-the-art scheduled schemes by at least $30\%$, even with imperfect feedback.
\end{abstract}


\glsresetall

\section{Introduction}

Goal-oriented communication is a new paradigm that aims at overcoming the limits of traditional communication systems by considering the \emph{meaning and purpose} of data, i.e., their value for a specific application~\cite{gunduz2023beyond}. Goal-oriented schemes consider the relevance of information, taking into account the shared context of the communicating agents, timing and bandwidth constraints, and the application-level performance metric that needs to be optimized. Research on the subject gained steam after the development of joint source-channel coding~\cite{bourtsoulatze2019deep} and has since been extended to wider \emph{semantic} aspects~\cite{shao2024theory}, is mostly focused on goal-oriented compression. Instead of classical reliability metrics, the semantic approach defines a complex, application-dependent distortion function: even if part of a message is lost, distorted, or omitted, the objective is to convey the intended meaning.

On the other hand, a parallel approach has been developed by the \gls{iot} community, focusing on \emph{medium access} instead of coding. In this case, the relevance of information depends on the error of a remote monitor that estimates the state of a dynamic process through sensor updates. The accuracy of the estimate will tend to degrade over time, unless new updates are received. \gls{aoi}, which represents the time elapsed since the generation of the last received status report~\cite{kaul2011minimizing}, captures this basic relation~\cite{kosta2017age}, but it is only a proxy for the actual relevance of sensory information, which depends on the stochastic evolution of the process. The \gls{voi} is a more recent metric that directly considers goal-oriented aspects by measuring the estimation error directly, allowing for more context-aware access schemes, but also increasing their complexity. In order to capture both the need for fresh updates and their relevance~\cite{lu2024semantics}, the \gls{aoii} considers a linear penalty counting the time elapsed since the last variation of system conditions \cite{maatouk2020age}.

The design of medium access schemes that can minimize \gls{aoi} or \gls{aoii} is an important problem in goal-oriented communication, as the relevance of sensor information is known to individual nodes, requiring a distributed approach. This is particularly relevant in scenarios with a large number of sensors and relatively rare events in each location, such as anomaly tracking~\cite{de2021survey}: scheduled schemes can minimize \gls{aoi}, or even the expected \gls{voi}~\cite{holm2023goal}, but the centralized scheduler cannot be aware of anomalies, leading to a higher \gls{aoii}. However, most of the relevant literature still considers centralized setups due to the need to coordinate transmissions~\cite{ayan2024optimal} to avoid the collision issue that plagues classical random access protocols such as ALOHA~\cite{badia2021impact}, 
even when using feedback from the common receiver~\cite{fahim2018towards} to 
resolve collisions by computing the state of other contending nodes~\cite{madueno2014efficient}.

The study of random access protocols that can act in a truly goal-oriented fashion, minimizing \gls{aoii} and fully exploiting the knowledge that centralized schemes lack, is still in its infancy~\cite{nayak2023decentralized}, as the analysis of \gls{aoii} is complex even for simple ALOHA-based protocols~\cite{yavascan2021analysis,munari2024monitoring,li2023timeliness}. This work aims at filling this gap by designing a distributed scheme that uses \gls{del}~\cite{baltag2004logics} to allow nodes to employ deductive reasoning over others' states based on \emph{common knowledge} information about their behavior. This can reduce both the frequency of collisions~\cite{maatouk2020age_csma} and the time needed to resolve them~\cite{pan2022age}.

We design \gls{delta}, a protocol that adopts \gls{del} to allow sensors to minimize \gls{aoii} distributedly. Each node considers its belief that it is the one with the highest \gls{aoii} and then acts accordingly: listening to acknowledgments guarantees that it is able to track everyone else's \gls{aoi}, using this information to update its belief over others' \gls{aoii}. The protocol considers a simple binary relevance model, which can however represent a variety of applications, such as \emph{(i)}, a set of wireless sensors reporting anomalies, e.g., excessive temperatures in a factory setting, to a common access point, in which the sensor detecting the occurrence of an anomaly remains in an alert state until it successfully reports it~\cite{abd2019role}, or \emph{(ii)}, a scenario in which agents request access to computing resources over a shared channel, sending a request/interrupt to the common computing engine~\cite{yang2012poll} when they receive a task~\cite{scheuvens2021state}.

To the best of our knowledge, we are the first to combine \gls{del} and goal-oriented communication, designing a random access protocol that exploits this information to provide superior performance over scheduled approaches. The contributions of this paper are listed as follows. 
\begin{itemize}
    \item We introduce \gls{delta}, a random access protocol based on inference reasoning, formally proving that it can allow multiple sensors to efficiently operate in a goal-oriented fashion based on common knowledge information;
    \item We analyze the protocol settings, providing an exact optimization framework for the collision resolution phase of the protocol and an approximate semi-Markov model for the epistemic reasoning phase;
    \item We provide an analysis of the effects of various feedback models, showing that the protocol is robust to errors in the feedback channel, degrading gracefully even in very difficult scenarios.
\end{itemize}
\gls{delta} can reduce the probability that the \gls{aoii} is over a set threshold by $30-80\%$ with respect to scheduled schemes if the offered load is below $0.5$, achieving much better performance than existing random access schemes. A preliminary version of this work was presented as a conference paper~\cite{chiariotti2025delta}. There are two major contributions in this work compared to~\cite{chiariotti2025delta}. First, we design a collision resolution scheme that is more advanced than the one in~\cite{chiariotti2025delta}, with a superior performance under ideal feedback. Second, we analyze the impact of imperfect feedback. Several feedback models are introduced for this purpose. The results confirm the robustness of \gls{delta} with respect to imperfect feedback.

The rest of this paper is organized as follows: first, Sec.~\ref{sec:rel} presents the state of the art. Sec.~\ref{sec:sys} then defines the communication system model, and the \gls{delta} protocol is specified in Sec.~\ref{sec:delta}, along with the theoretical analysis of its parameters. We then describe the simulation results in Sec.~\ref{sec:sim}, while Sec.~\ref{sec:conc} concludes the paper and presents some possible avenues of future work.

\section{Related Work}\label{sec:rel}

The analysis of \gls{aoii} and other \gls{aoi} extensions
in distributed settings is still in its infancy. 
The existing random access schemes that target information freshness, either require a certain side coordination, or a traffic is extremely sporadic \cite{li2023timeliness,yates2021age}. Even though it was studied in the seminal paper that first defined \gls{aoi}~\cite{kaul2011minimizing}, where the metric was originally introduced for vehicular networks, relatively few works have explicitly considered medium access. A common approach is to treat centralized coordinated access~\cite{bedewy2021optimal,holm2023goal}, due to the complexity of keeping track of the system state in distributed schemes, as well as information locality: since sensors operate without knowing what the others measure, the collision risk becomes acute unless access is centrally scheduled. Several recent studies~\cite{munari2021information} considering \gls{aoi} in random access channels point out how collisions have a detrimental effect on \gls{aoi}, even when considering carrier sensing~\cite{maatouk2020age_csma} and collision resolution mechanisms~\cite{pan2022age}. The efforts to prevent nodes from entering collisions are mostly circumscribed to the threshold ALOHA approach~\cite{yavascan2021analysis}, which can be adapted dynamically to time-varying traffic conditions~\cite{moradian2024age}. However, threshold-based methods can be efficient for \gls{aoi} but are suboptimal for anomaly reporting due to the overhead incurred due to waiting until an \gls{aoii} threshold is reached \cite{de2023unified}. 

Deterministic access quickly becomes \gls{aoi}-optimal for large networks \cite{jiang2019timely}; however, this only holds if the traffic is intense. There are very few investigations on the freshness of anomaly reporting, which is not expected to be persistent. Most anomaly tracking applications, where staleness is better quantified by \gls{aoii}, do not require constant updates and avoid unnecessary transmissions, improving battery lifetime and congestion~\cite{holm2023goal}. Scenarios include vehicular flow management in which critical reporting by a vehicle is not constant and depends on its position~\cite{thota2020v2v}, environmental supervision in smart agriculture, wildlife tracking, or monitoring for safety and security purposes in domotic, industrial, or smart grid scenarios~\cite{kontar2021internet}. Even medical supervision of elderly or chronic patients likely only reports relevant condition changes \cite{ning2021mobile}. In all these scenarios the traffic is intermittent, but far from sporadic (e.g., vehicular communications may require an exchange of data with an update every second or so \cite{uhlemann2016connected}), and the tracked anomalies are sudden and variable across the users. In this context, analyzing \gls{aoii} in more complex reservation-based protocols is often only possible as the number of nodes grows to infinity~\cite{hui2023fresh}, while precise results for finite networks have been provided just for simple schemes, such as ALOHA~\cite{cocco2023remote}. To the best of our knowledge, the only work to actively optimize \gls{aoii} instead of analyzing existing schemes is~\cite{nayak2023decentralized}, whose results are still inferior to simple round-robin. 

We then consider the work on epistemic logic, a branch of formal reasoning dealing with the inference, transfer, and update of knowledge among multiple agents~\cite{malandrino2023efficient,huang2024ai}. When knowledge evolves over time and successive interactions, this is referred to as \gls{del}, and finds applications in social networks and cryptography \cite{chen2019analysis}.
The solution is often obtained through meta-reasoning on whether \emph{other} agents are able to solve the problem. For example, in the well-known ``muddy children puzzle,'' agents may possess an individual trait (i.e., a dirty face) or not. This information is not directly available, as each agent only knows if others have the trait, and that at least one child does~\cite{kline2013evaluations}. Proceeding by induction, one can determine the exact number of muddy faces over a few rounds. 

There have been a few attempts at introducing \gls{del} at the network level, mostly driven by the use of AI-empowered devices. For example, \cite{kontar2021internet} discusses the ability of \gls{iot} systems to combine local knowledge of individual nodes through automated reasoning, so as to gain further meta-information. Quite recently,~\cite{huang2024ai} has explored AI for network virtualization, and leverages epistemic logic to improve over the uncertainties of AI with respect to traditional software-based virtual network functions. However, none of these or other similar proposals consider \gls{del} for medium access.

\section{System Model}\label{sec:sys}
Consider a discrete-time system with a set $\mc{N}$ of sensors (also referred to as nodes), each of which measures an independent quantity and can detect anomalies. We denote the number of nodes as $N=|\mc{N}|$ and the state at time step $t$ as $\mb{x}_t\in\{0,1\}^N$, whose $n$-th component $x_{n,t}$ corresponds to the state of sensor $n$ at time $t$. At any time slot, sensor $n$ may switch from the normal state $0$ to the anomalous state $1$ with probability $\lambda_n$. On the other hand, state $1$ is absorbing, i.e., the anomaly persists until the sensor successfully transmits a warning to the gateway. The transition matrix $\mb{A}_n$ is then
\begin{equation}
\mb{A}_n=\begin{pmatrix}
    1-\lambda_n & \lambda_n\\
    s_{n,t} & 1-s_{n,t}
\end{pmatrix},
\end{equation}
where $s_{n,t}\in\{0,1\}$ is an indicator variable which is equal to $1$ if $n$ successfully transmits at time $t$ and $0$ otherwise.\footnote{For the sake of simplicity, we consider transmissions to be instantaneous. The case in which transmissions incur a delay of $1$ slot can be dealt with by adding $1$ to all \gls{aoi} and \gls{aoii} measurements in the following.}
We then define the \gls{aoi} of node $n$ at time $t$, denoted as $\Delta_{n,t}$, as
\begin{equation}
    \Delta_{n,t}=t-\max_{\tau\in\{1,\ldots,t\}}\tau\, s_{n,t-\tau}.
\end{equation}
However, \gls{aoi} is not meaningful in our case, as a sensor might spend a long time with nothing to report: as long as its state is normal, new updates from it are not necessary. We then introduce the \gls{aoii} $\Theta_{n,t}$~\cite{maatouk2020age}, which is defined as
\begin{equation}
\Theta_{n,t}=t-\argmax_{\theta\in\{t-\Delta_{n,t}+1,\ldots,t\}}\theta\, x_{n,t-\theta}.
\end{equation}
As Fig.~\ref{fig:aoi_aoii} shows, the \gls{aoi} grows even while in the normal state, while the \gls{aoii} only grows in the anomalous state.

\begin{figure}
    \centering
\begin{tikzpicture}

\path [fill=green!30!white,draw] (0,1.9)--(0.5,1.9)--(0.5,2.2)--(1,2.2)--(1,2.5)--(1.5,2.5)--(1.5,2.8)--(2,2.8)--(2,1)--(0,1)--cycle;

\path [fill=green!30!white,draw] (2.5,1.3)--(3,1.3)--(3,1.6)--(3.5,1.6)--(3.5,1.9)--(4,1.9)--(4,2.2)--(4.5,2.2)--(4.5,2.5)--(5,2.5)--(5,1)--(2.5,1)--cycle;

\path [fill=green!30!white] (5.5,1.3)--(6,1.3)--(6,1.6)--(6.5,1.6)--(6.5,1.9)--(7,1.9)--(7,2.2)--(7.5,2.2)--(7.5,1)--(5.5,1)--cycle;

\path [fill=blue!30!white,draw] (0.5,1)--(0.5,1.3)--(1,1.3)--(1,1.6)--(1.5,1.6)--(1.5,1.9)--(2,1.9)--(2,1)--cycle;

\path [fill=blue!30!white,draw] (4,1)--(4,1.3)--(4.5,1.3)--(4.5,1.6)--(5,1.6)--(5,1)--cycle;

\path [fill=red!30!white,draw] (0.5,1)--(0.5,1.3)--(2,1.3)--(2,1)--cycle;

\path [fill=red!30!white,draw] (4,1)--(4,1.3)--(5,1.3)--(5,1)--cycle;

\path [draw] (5.5,1)--(5.5,1.3)--(6,1.3)--(6,1.6)--(6.5,1.6)--(6.5,1.9)--(7,1.9)--(7,2.2)--(7.5,2.2);

\draw[->] (0,1) --  node [above=0.1cm,very near end]{\footnotesize Time} (7.5,1);
\draw[->] (0,1) -- (0,3);
\foreach \i in {1,...,15}
    \draw[-] (0.5*\i-0.5,0.9)  --node[below=0.1cm]{\footnotesize$\i$} (0.5*\i-0.5,1);
\foreach \i in {0,...,6}
    \draw[-] (-0.1,0.3*\i+1) --node[left=0.1cm]{\footnotesize$\i$} (0,0.3*\i+1);

\path [fill=green!30!white,draw] (5.3,3.1)--(5.3,2.8)--(5.7,2.8)--(5.7,3.1)--cycle;

\path [fill=blue!30!white,draw] (5.3,2.4)--(5.3,2.7)--(5.7,2.7)--(5.7,2.4)--cycle;

\path [fill=red!30!white,draw] (5.3,2)--(5.3,2.3)--(5.7,2.3)--(5.7,2)--cycle;

\path [draw] (5.2,1.9)--(5.2,3.2)--(6.45,3.2)--(6.45,1.9)--cycle;

\node[anchor=west] at (5.7,2.15){\scriptsize State};
\node[anchor=west] at (5.7,2.55){\scriptsize AoII};
\node[anchor=west] at (5.7,2.95){\scriptsize AoI};

\end{tikzpicture}\vspace{-0.7cm}
    \caption{Example of the AoI and AoII evolution for a node.}\vspace{-0.4cm}
    \label{fig:aoi_aoii}
\end{figure}

We consider the wireless communication system to operate in \gls{tdd} mode, so that each time slot is divided in an uplink and downlink part. During the uplink part, each sensor may transmit or remain silent. The uplink is modeled as a collision channel, in which transmissions are never successful if more than one node is active. If a single node $n$ transmits, its packet erasure probability is $\varepsilon_n$.
During the downlink part, all sensors are in listening mode. If the uplink transmission was successful, the acknowledgment (ACK) packet from the gateway informs all nodes of the identity of the transmitter, while if it was unsuccessful, either because of a collision or a wireless channel erasure, a \gls{nack} packet informs all nodes of the failure, but does not report the identity of the transmitting nodes. Finally, if no node transmitted, the gateway is silent~\cite{badia2009effect}.

We will consider four different models for the ACK and \gls{nack} transmission channel from the gateway to the nodes:
\begin{itemize}
    \item An \emph{ideal} feedback channel, in which all nodes receive the messages without errors;
    \item A \emph{noisy} feedback channel, in which ACKs and \glspl{nack} are always distinguished, but the decoded identity of the intended recipient of the ACK is a Gaussian random variable with a standard deviation $\sigma_f$, as explained below;
    \item An \emph{erasure} feedback channel, in which each node may be unable to decode the ACK with probability $\varepsilon_f$, but knows whether a feedback message was sent;
    \item A \emph{deletion} feedback channel, in which a node is unable to even know if a feedback message was transmitted or not with probability $\omega_f$.
\end{itemize}
In general, the protocol is robust to an imperfect feedback channel, and we will discuss the countermeasures to deal with this case in the following. The noisy model is inspired by new \gls{iot} technologies such as wake-up radio: if acknowledgments use extremely simple analog encoding (e.g., by encoding node identifiers as the duration of a signal), the electronics implementing the receiver can be designed to consume orders of magnitude less than a standard radio. In this case, confusing ACKs and \glspl{nack} becomes almost impossible, as the code can be designed for a wide separation of the two, but the duration of the ACK signal may be misinterpreted by nodes, leading to a certain probability of error over the node ID. In this case, we consider a Gaussian noise over the decoded ID, $w\sim\mathcal{N}(0,\sigma_f^2)$: if node $n$ receives an ACK for a packet sent by node $m$, the decoded ID is
\begin{equation}
    \hat{m}_n=\Mod(\Int(m-1+w), n)+1,
\end{equation}
where $\Mod(m,n)$ is the integer modulo function.

Finally, if node $n$ transmitted during the slot, it will always assume that an ACK is meant for its own packet independently of the noise, as only one packet can be acknowledged in a given slot. On the other hand, the erasure and deletion models correspond to more classical digital feedback channel models, in which the nodes are in receive mode during the downlink phase of each round. This usually ensures a very low feedback error probability, as the gateway can transmit using a high power and a robust modulation and coding, but requires a higher energy expense for the nodes.

In the following, we will refer to random variables using capital letters, e.g., $X$, while their realizations will use the corresponding lowercase letter, e.g., $x$. The \gls{pmf} of $X$ will be indicated as $p_X(x)$, and the corresponding \gls{cdf} will be $P_X(x)$. Vectors are indicated as bold lowercase letters, e.g., $\mb{x}$, whose $n$-th element is denoted by $x_n$. Matrix symbols are bold capital letters, e.g, $\mb{A}$, whose $m,n$-th element is denoted by $A_{m,n}$.

\section{The DELTA Protocol}\label{sec:delta}

Distributed protocols that can take the content of sensor observations into account are rare in the relevant literature: while a centralized controller cannot exploit the knowledge of the sensors' true observations, distributed protocols are often plagued by collisions \cite{yavascan2021analysis,maatouk2020age_csma,li2023timeliness}. Sensors can decide whether and when to transmit based on their own observations, but they do not know what other sensors are observing, and which decisions they might make as a result. This often causes inefficiencies that have made distributed protocols valuable only for niche applications: to reduce the risk of collisions, sensors need to randomly abstain from transmitting, increasing their \gls{aoii} even when there would be no need to do so.

The \acrfull{delta} protocol
is based on the notion of \emph{common knowledge} as defined in~\cite{baltag2004logics}. \gls{del} is a formal framework to describe the dynamics of beliefs in multi-agent systems, which distinguishes between general and common knowledge proposition. A proposition is general knowledge if its truth value is known to all agents, while for it to be common knowledge, the fact that it is general knowledge also needs to be known to all agents, extending recursively to infinity. The use of common knowledge-based Bayesian reasoning allows \gls{delta} nodes to maintain a shared understanding of the state of the system, which each sensor can combine with its own private observations to make communication decisions. Furthermore, the public outcome of these decisions can be used by sensors to infer other nodes' private knowledge, following a Bayesian framework. 
The crucial aspect to enable this is the public nature of ACKs. In the following, we will only consider the ideal and noisy feedback channel cases, but we will discuss how to adapt \gls{delta} to an imperfect feedback channel in Sec.~\ref{sec:delta-imperfect}.

\subsection{Protocol Definition and Correctness}

The \gls{delta} protocol includes $4$ phases, and transitions between them only depend on publicly available information, e.g., the outcome of the previous slot.

The \emph{\gls{zw}} phase is the normal state of operation: during this phase, each sensor transmits whenever its state changes, i.e., an anomaly occurs. This allows us to keep the \gls{aoii} equal to $0$ when the system is empty. Sensors remain in this phase until a transmission fails due to multiple sensors simultaneously observing anomalies or a wireless channel erasure. 
As the gateway transmits a NACK signal to inform sensors of the collision, all sensors switch to the \emph{\gls{cr}} phase \cite{pan2022age}, recording their membership in the collision set through an indicator variable $m_{n,t}$.

\begin{lemma}\label{lm:zw}
Under an ideal or noisy feedback channel, as long as the system remains in the \gls{zw} phase, all sensors are in state $0$, and the state is common knowledge.
\end{lemma}
\begin{IEEEproof}
Let us consider slot $t$, knowing that all sensors are in state $0$ at time $t-1$. Since nodes in state $1$ always transmit during phase \gls{zw}, a silent slot, in which case nobody had anything to transmit, can be interpreted by all nodes as the state remaining the same~\cite{plaza2007logics}. In formal terms, $\Theta_n=0$ is a precondition for a node remaining silent. The same holds for a successful transmission, i.e., a single node transmitting and resetting its \gls{aoii} and state to $0$. On the other hand, a NACK may be caused by a wireless channel loss or a collision between multiple transmitters. In this case, all nodes move to the \gls{cr} phase. Under ideal or noisy feedback, all nodes know whether the feedback was an ACK or a \gls{nack}, and this is common knowledge. The phase of the protocol, and the state of the system, are then also common knowledge.
\end{IEEEproof}

During the \gls{cr} phase, nodes with $m_{n,t}=0$ never transmit. In the first slot after the collision, members of the collision set transmit with a certain probability $p$. In the following slots, the nodes keep transmitting with the same probability until there is a successful transmission, i.e., an ACK is received: in this case, the nodes transition to the \emph{\gls{ce}} phase. During this phase, nodes that are not in the collision set remain silent, while the node that successfully transmitted exits the collision set by setting $m_{n,t}=0$. All remaining members of the collision set transmit with probability $1$. This strategy increases the resolution time if there are more than $2$ colliding nodes, as it causes another collision, but this case is relatively rare due to the low traffic, and it confers a major advantage: the second collision allows all nodes to know that the initial collision is still unresolved, and that there should be another \gls{cr} phase. Conversely, successful or silent slots only happen when the collision set becomes empty, and nodes can safely switch from the \gls{ce} to the \emph{\gls{bt}} phase.

\begin{lemma}\label{lm:cr}
The switches between phases \gls{cr}, \gls{ce}, and \gls{bt} are common knowledge under the ideal and noisy feedback channel models.
\end{lemma}
\begin{IEEEproof}
After the switch from \gls{zw} to \gls{cr}, state $\mb{x}_t$ is not common knowledge any more: each node knows its own state and \gls{aoii}, but not others'. However, we can use public announcements to infer phase changes: if a transmission in the \gls{cr} phase is successful, the transmitting node \emph{was} part of the collision set, but its state is reset to $0$, and the system switches to \gls{ce}. The reception of an ACK in the \gls{cr} phase then triggers to switch to the \gls{ce} phase, and we note that ACKs are received by every sensor. We can then use the precondition on outcomes in the \gls{ce} phase: as all remaining members of the collision set transmit, we know that the set is non-empty only after a \gls{nack}, which represents a public announcement of a switch back to \gls{cr}. The next phase is then common knowledge. If we consider the noisy feedback model, the proof is still valid, as the identity of the node whose packet is being acknowledged might be mistaken, but ACKs, NACKs, and silent slots can always be distinguished perfectly.
\end{IEEEproof}

Finally, the \gls{bt} phase allows sensors to gradually go back to normal: as the sequence of \gls{cr} and \gls{ce} phases can take several steps, anomalies may have accumulated, and several sensors may have a high \gls{aoii}. Consequently, the sensors need to get back to a state in which they have common knowledge that everyone is in state $0$ before \gls{zw} operation can safely resume.

\begin{figure}
    \centering
\begin{tikzpicture}[->, shorten >=2pt, line width=0.5 pt, node distance =1 cm]
                        ]
\node (bt)  [circle,draw,minimum height=1cm] {BT};
\path (bt) edge [loop left] node [left]{\small not NACK} (bt);

\node (zw)  [circle,draw,minimum height=1cm,above=1cm of bt] {ZW};
\path (zw) edge [loop left] node [left]{\small not NACK} (zw);

\node (cr)  [circle,draw,minimum height=1cm,right=2.5cm of zw] {CR};
\path (cr) edge [loop right] node [right]{\small not ACK} (cr);

\node (ce)  [circle,draw,minimum height=1cm,right=2.5cm of bt] {CE};

\path (cr) edge  [bend left] node [right]{\small ACK} (ce);
\path (ce) edge  [bend left] node [left]{\small NACK} (cr);

\path (ce) edge  node [below]{\small not NACK} (bt);

\path (bt) edge  [bend left] node [left]{\small $\max \bm{\varphi}_t=0$} (zw);
\path (zw) edge  node [above]{\small NACK} (cr);

\path (bt) edge  node [above left]{\small NACK} (cr);




\end{tikzpicture}\vspace{-0.7cm}
    \caption{DELTA state diagram.}\vspace{-0.4cm}
    \label{fig:protocol_state}
\end{figure}

Let us denote the highest possible \gls{aoii} that a node might have given the common knowledge information as $\psi_{n,t}$. By definition, $\Theta_{n,t}\leq\psi_{n,t} \quad \forall t,n$.
Node $n$'s \gls{aoii} $\Theta_{n,t}$ is the highest if no node has higher \gls{aoii}, and the activation of each node is independent.
The probability that node $n$ has the highest \gls{aoii}, given the vector $\bm{\psi}_t$, is then
\begin{equation}\label{eq:bt_prob}
f_{n,t}\left(\Theta_{n,t},\bm{\psi}_t\right)=\prod_{m\neq n}\left(1-\lambda_m\right)^{\left[\psi_{m,t}-\Theta_{n,t}+1\right]^+},
\end{equation}
where $[x]^+=\max(0,x)$ is the positive part operator.
In the \gls{bt} phase, we set a threshold $F$, and node $n$ transmits with probability $1$ if $f_{n,t}>F$. If $\bm{\psi}_t=\mb{0}_N$, i.e., the all-zero vector of length $N$, the system goes back to the \gls{zw} phase. The \gls{delta} phase diagram is shown in Fig.~\ref{fig:protocol_state}.

\begin{theorem}\label{th:full_knowledge}
The protocol phase and $\bm{\psi}_t$ are always common knowledge if the feedback channel is ideal.
\end{theorem}
\begin{IEEEproof}
We have $\psi_{n,t}=0\ \forall n\in\mc{N}$ during the \gls{zw} phase as a direct consequence of Lemma~\ref{lm:zw}. If we consider the sequence of \gls{cr} and \gls{ce} phases starting at time $t$ from phase \gls{zw} and ending after $k$ slots, there are two common knowledge propositions: firstly, as stated in Lemma~\ref{lm:cr}, switches between phases are common knowledge. Secondly, it is common knowledge that nodes outside the collision set were in state $0$ at time $t$, as they were in the \gls{zw} phase and did not transmit.

The nodes with an \gls{aoi} lower than $j$ were in the collision set, and their transmissions reset their state to $0$: their \gls{aoii} is capped to their \gls{aoi} by definition. When the \gls{bt} phase begins,
\begin{equation}
\psi_{n,t+k}=\min\left(k,\Delta_{n,t}\right),\ \forall n\in\mc{N}.
\end{equation}
During the \gls{bt} phase, communication decisions are based on the probability defined in~\eqref{eq:bt_prob}. The outcome of each slot is then broadcasted: if sensor $n$ did not transmit at time $t$,
\begin{equation}\label{eq:adj_age}
\psi_{n,t+1}=1+\sup
\left(\tilde{\theta}\in\{0,\ldots,\psi_{n,t}\}: f_{n,t}\left(\tilde{\theta},\bm{\psi}_{t}\right)<F\right).
\end{equation}
If the outcome was silence or a successful transmission, all nodes (except the successful one, whose \gls{aoii} was reset to $0$) were silent. On the other hand, if the outcome of the round was a collision,  all nodes except the members of the collision set were silent, by definition. The value of $\psi_{n,t+1}$ can then safely be reset for all nodes, as all colliding nodes will transmit again before the next \gls{bt} phase. During subsequent collision resolution cycles, $\psi_{n,t}$ increases by the duration of the cycle, and is reset to $0$ for nodes that successfully transmit. The return to phase \gls{zw} depends only on $\bm{\psi}_t$. On the other hand, if the feedback channel is imperfect, the nodes may switch to phase \gls{zw} at different times, based on their (correct or incorrect) beliefs on other nodes' maximum possible \gls{aoii}.
\end{IEEEproof}

We note that collisions are more common in the \gls{bt} phase than in \gls{zw}, as nodes must be more aggressive to gradually reduce $\bm{\psi}_t$. All collisions are handled identically, regardless of the phase during which they originated. The full decision-making algorithm for each sensor is presented as Alg.~\ref{alg:full_scheme}.

\begin{figure}[t]
\vspace{-8pt}
\begin{algorithm}[H]
\caption{Pseudocode of the DELTA protocol}
\label{alg:full_scheme}
\begin{algorithmic}[1]
\small

\Require phase, $F$, $\mb{p}$, $x_{n,t}$, NACK, $m_{n,t}$, $c_t$, $\bm{\psi}_{t-1}$
\If{NACK}
    \If{$\text{phase}=\text{CE}$}
        \State $c_t\gets c_t+1$
    \EndIf
    \State $\text{phase}\gets\text{CR}$
\EndIf
\If{ACK \textbf{and} $\text{phase}=\text{CR}$}
    \State $\text{phase}\gets\text{CE}$
\EndIf
\If{$\text{phase}=\text{BT}$}
    \State $\bm{\psi}_t\gets$\Call{UpdateMaximumPossibleAoII}{$\bm{\psi}_{t-1}$} 
    \If{$\max(\bm{\psi}_t)=0$}
        \State $\text{phase}\gets\text{ZW}$
    \EndIf
\EndIf
\If{$\text{phase}=\text{CE}$ \textbf{and} (not NACK)}
    \State $\text{phase}\gets\text{BT}$, $c_t\gets 0$
\EndIf
\If{$x_{n,t}=0$}
    \State \Return{$0$}
\Else
    \Switch{phase}
        \Case{ZW: \Return{$1$}}
        \EndCase
        \Case{CR: \Return{$m_{n,t}\times$\Call{BernoulliSample}{$p(c_t)$}}}
        \EndCase
        \Case{CE: \Return{$m_{n,t}$}}
        \EndCase
        \Case{BT: \Return{\Call{HighestAoIIProb}{$\theta_t,\bm{\psi}_t$}$>F$}}
        \EndCase
    \EndSwitch
\EndIf
\end{algorithmic}
\end{algorithm}\vspace{-0.8cm}
\end{figure}


\subsection{Collision Resolution Phase Optimization}

The expected number of slots $\tau_c$ required to resolve a collision depends on the number $C$ of colliding nodes, which transmit with the same probability $p$ until the collision is resolved. The probability of success in any given slot when there are $c$ colliding nodes is
\begin{equation}
\sigma(c,p,\varepsilon)=(1-\varepsilon_n)\Bin(1;c,p),
\end{equation}
where $\Bin(k;N,p)=\binom{N}{k}p^k(1-p)^{N-k}$ is the binomial \gls{pmf}. After the first ACK, the remaining colliding nodes transmit with probability $1$ in the \gls{ce} phase. This means that $C-1$ nodes will collide if $C>2$. We then define vector $\mb{p}$, whose $i$-th element represents the transmission probability in the $i$-th collision resolution phase. 

If all nodes have the same $\varepsilon$, we can represent the cycle starting from $c$ colliding nodes as an absorbing Markov chain with $c$ states, representing each individual \gls{cr} phase. The transition from one state to the next is the \gls{ce} phase, and the structure of the protocol prevents the size of the collision set from increasing. The transition probability matrix is
\begin{equation}
\mb{P}_c=\begin{pmatrix}
\mb{B} & \sigma(2,c-1)\mb{u}^{c-1}_{c-1}\\
(\mb{0}_{c-1})\T & 1
\end{pmatrix},
\end{equation}
where $\mb{u}^n_N$ is identical to $\mb{0}_N$ except for element $n$, which is equal to $1$, and the elements of matrix $\mb{B}$ are\footnote{In the following transition matrices, we omit transitions with probability $0$ for the sake of brevity.}
\begin{equation}
B_{ij}=\begin{cases}
1-\sigma(c-i+1,p_i,\varepsilon), &j=i;\\
\sigma(c-i+1,p_i,\varepsilon), &j=i+1.
\end{cases}
\end{equation}
The time $\tau_c$ until absorption, i.e., until the collision is fully resolved, follows a discrete phase-type distribution characterized by the matrix $\mb{P}_c$. The \gls{cdf} of $\tau_c$ is simply given by the corresponding element of the $t$-step matrix, $p_{\tau_c}(t)=\left(\mb{P}_c\right)^t_{{1,c}}$.
In the case where $c=1$, i.e., when a single node's transmission failed because of the channel, the time until absorption reduces to a geometric random variable, i.e., $\tau_1\sim\Geo(p_1)$.

\begin{theorem}\label{th:expected_cr}
If the colliding set was a singleton, i.e., $C=1$, the expected duration of the subsequent \gls{cr} and \gls{ce} cycle is
\begin{equation}
\E{\tau_1}=1+\big((1-\varepsilon)p_1\big)^{-1}.
\end{equation}
For a set of $c>1$ colliding nodes with the same $\varepsilon$, the expected duration of a cycle of \gls{cr}-\gls{ce} phases, which begins after the initial collision and ends when the collision set is empty, is
\begin{equation}
\E{\tau_c}=c-1+\varepsilon+\frac{\varepsilon}{(1-\varepsilon)p_c}+\!\sum_{i=0}^{c-2}\frac{1}{\sigma(c-i,p_{i+1},\varepsilon)}.
\end{equation}

\end{theorem}
\begin{IEEEproof}
We begin by proving the theorem in the singleton case, in which there is a single \gls{cr} phase, whose duration is geometrically distributed with parameter $(1-\varepsilon)p_1$. An additional slot needs to be added to account for the \gls{ce} phase.

In the general case, the expected time until absorption of a Markov chain is hard to compute, but the structure of the transition matrix simplifies the problem. Any state $i$ is reached from $i-1$ with a successful transmission after a geometrically distributed number of failures, i.e., self-transitions:
\begin{equation}
\E{\tau_{i-1,i}|C=c,p_{i-1}}=\big(\sigma(c-i,p_{i-1},\varepsilon)\big)^{-1}.
\end{equation}
The number of self-transitions in each state is independent from what happens in other states due to the Markov property, and the protocol requires $c-1$ \gls{cr} phases to reach the absorbing state $c$. Additionally, there are $c-2$ collisions caused by the intermediate \gls{ce} phases, during which the nodes discover that the collision set is not empty. Finally, we have one more \gls{ce} phase from the last colliding node when we have reached state $c$. If the transmission is successful, the cycle is over, but if there is a wireless channel loss, we have one more singleton collision resolution cycle after it.
\end{IEEEproof}

However, the value of $C$ is unknown to the sensors. If we consider the \gls{zw} phase in a system in which all sensors have the same activation probability $\lambda$, we get
\begin{equation}
p_{C}(c|\text{ZW})=
    \Bin(c;N,\lambda)\left[1-(1-\varepsilon)\delta(c,1)\right],
\end{equation}
where $\delta(m,n)$ is the Kronecker delta function, equal to $1$ if the two arguments are equal and $0$ otherwise.
We can also easily get the total failure probability $p_f(\text{ZW})=\sum_{c=1}^N p_{C}(c|\text{ZW})$.
We can then apply the law of total probability, adding the $c-1$ \gls{ce} phases as in Theorem~\ref{th:expected_cr}, to obtain the \gls{cdf} of the duration of a collision resolution cycle:
\begin{equation}\label{eq:cr_duration}
\begin{aligned}
P_{\tau}(t|\text{ZW})=\frac{\varepsilon(1{-}\varepsilon)}{p_f(\text{ZW})}\bigg[\!\Bin(1;N,\lambda)\left(1-\eta_1^{t-1}\right)
\\+\!\sum_{c=2}^{\mathclap{\min(N,t)}}\Bin(c;N,\lambda)\bigg(\!\frac{\left(\mb{P}_c\right)^{t{-}c{+}1}_{{1,c}}}{\varepsilon}+\sum_{k=1}^{\mathclap{t-2c+1}}\left(\mb{P}_c\right)_{{1,c}}^{t{-}c{-}k}\eta_c^{k{-}1}p_c\bigg)\!\bigg],
\end{aligned}
\end{equation}
where $\eta_c=1-(1-\varepsilon)p_c$.

\begin{theorem}\label{th:collision_opt}
There is a single optimal transmission probability $\mb{p}^*$ that minimizes the expected duration
\begin{equation}\label{eq:delay_minimization}
    \mb{p^*}=\argmin_{\mb{p}\in(0,1)^N}\sum_{c=1}^N p_C(c|\text{ZW})\E{\tau_c},
\end{equation}
if all nodes have the same $\lambda$ and $\varepsilon$, and $p^*_i$ is the solution of
\begin{equation}\label{eq:minimization_derivative}
\frac{\Bin(1;N_i,\lambda)\varepsilon}{(p_i^*)^2}+\sum_{c=2}^{N_i}\Bin(c;N_i,\lambda)\frac{1-cp_i^*}{c(p_i^*)^2(1-p_i^*)^c}=0,
\end{equation}
where $N_i=N-i+1$. In the $N$-th \gls{cr} phase, $p_N^*=1$.
\end{theorem}
\begin{IEEEproof}
Since each \gls{cr} phase is independent from all others, we can optimize each element of $\mb{p}$ separately to minimize the expected duration. We then take the first probability:
\begin{equation}
\begin{aligned}
p_1^*=&\argmax_{p\in(0,1)}\left[\sum_{c=1}^N \frac{p_C(c|\text{ZW})(1-(1-\varepsilon)\delta(c,1))}{p_f(\text{ZW})\sigma(c,p,\varepsilon)}\right]\\
=&\argmax_{p\in(0,1)}\left[\sum_{c=1}^N w_c \frac{1-\varepsilon}{cp(1-p)^{c-1}}\right].
\end{aligned}
\end{equation}
In order to prove that it is convex, we only need to prove that each individual component is convex. The first one, with $c=1$, is proportional to $p^{-1}$, so it is convex for $p>0$. We show that components with $c>1$ are also convex by taking the second derivative of $(\sigma(c,p,\varepsilon))^{-1}$ with respect to $p$:
\begin{equation}
    \frac{\partial^2 \big(\sigma(c,p,\varepsilon)\big)^{-1}}{\partial p^2}=\frac{c(c+1)p^2-2(c+1)p+2}{(1-\varepsilon)cp^3(1-p)^{c+1}}.
\end{equation}
As $c>1$, $p\in(0,1)$, and $1-\varepsilon$ is always positive, so is the denominator. The second derivative is then positive if
\begin{equation}
    c(c+1)p^2-2(c+1)p+2>0.
\end{equation}
This quadratic equation has no real solution for $c>1$.
We can trivially prove that the two extremes, $p=0$ and $p=1$, lead to an infinite expected duration for $N>1$: if $p=0$, no node ever transmits, while if $p=1$, the nodes will keep colliding forever whenever the remaining collision set is not a singleton \cite{badia2021impact}. The maximum is then inside the interval for $N>1$.

Finally, we can prove that~\eqref{eq:minimization_derivative} is a multiple of the first derivative of the optimization function in~\eqref{eq:delay_minimization}, and finding its root in $(0,1)$ is equivalent to finding the minimum. As the solution of~\eqref{eq:minimization_derivative} involves a hypergeometric function, there is no closed-form solution, but it can be approximated efficiently with the bisection method and stored in a look-up table.
\end{IEEEproof}

\subsection{DELTA+}
A fixed transmission probability still does not fully account for the information received through public announcements: each failed or silent slot can be used as a Bayesian update. This principle was adopted as part of the Sift protocol~\cite{tay2004collision}, which provided an optimal solution for a known number of colliders and an approximated one with an unknown number. In our case, we 
the initial distribution of the number of colliders in the first \gls{cr} phase is 
\begin{equation}
    \phi_0(c)=\mathbbm{1}(c-1)\frac{\big(1-(1-\varepsilon)\delta(c,1)\big)\Bin(1;N,\lambda)}{\varepsilon\Bin(1;N,\lambda)+\sum_{c'=2}^N\Bin(c';N,\lambda)},
\end{equation}
where $\mathbbm{1}(x)$ is the stepwise function, equal to $1$ if $x\geq0$ and $0$ otherwise.
 We can then update the belief distribution after an ACK by applying Bayes' theorem:
\begin{equation}
    \phi^{\text{CR}}_{j+1}(c|\text{ACK})=\frac{\phi_{j}(c+1)\left[(c+1)p_j(1-\varepsilon)(1-p_j)^{c}\right]}{\sum_{c'=1}^{N}\phi_{j}(c')\left[c'p_j(1-\varepsilon)(1-p_j)^{c'-1}\right]}.
\end{equation}
After a silent slot, we get
\begin{equation}
    \phi^{\text{CR}}_{j+1}(c|\text{SIL})=\frac{\phi_{j}(c)(1-p_j)^c}{\sum_{c'=0}^{N}\phi_{i,j}(c')(1-p_j)^{c'}}.
\end{equation}
Finally, we can perform a similar update after a NACK:
\begin{equation}
    \phi^{\text{CR}}_{j+1}(c|\text{NACK})=\frac{\phi_{j}(c)p_{\text{NACK}}(c)}{\sum_{c'=0}^{N}\phi_{j}(c')p_{\text{NACK}}(c')},
\end{equation}
where $p_{\text{NACK}}(c)$ is
\begin{equation}
    p_{\text{NACK}}(c)=1-(1-p_j)^c-cp_j(1-\varepsilon)(1-p_j)^{c-1}.
\end{equation}
After an unsuccessful \gls{ce} phase, we update the belief as
\begin{equation}
    \phi^{\text{CE}}_{j+1}(c|\text{NACK})=\frac{\phi_{j}(c)\mathbbm{1}(c-1)\big(1-(1-\varepsilon)\delta(c,1)\big)}{\varepsilon\phi_{j}(1)+\sum_{c'=2}^{N}\phi_{j}(c')}.
\end{equation}
Using this belief distribution, the optimal transmission probability $p_j^*$ is the solution of
\begin{equation}
    \frac{\phi_j(1)}{(p_j^*)^2}+\sum_{c=2}^N\frac{(1-cp_j^*)\phi_j(c)}{c(p_j^*)^2(1-p_j^*)^c}=0.
\end{equation}
The proof that this solution is optimal trivially follows from Theorem~\ref{th:collision_opt}. We will refer to the version of the protocol using this slot-level belief update as \gls{delta}+, to distinguish it from the basic version, which is computationally much lighter (probabilities can be stored as a look-up table) but also expected to perform slightly worse due to the slower collision resolution process.

\begin{figure}
    \centering
\begin{tikzpicture}[->, shorten >=2pt, line width=0.5 pt, node distance =1 cm]
                        ]

\node (cr0) [circle,draw,minimum height=1.1cm] {\scriptsize CR$(0)$};
\node (cr1) [circle,draw,minimum height=1.1cm,right=0.75cm of cr0] {\scriptsize CR$(1)$};
\node (cr2)  [circle,draw,minimum height=1.1cm,right=0.75cm of cr1] {\scriptsize CR$(2)$};
\node (cr3)  [circle,draw,minimum height=1.1cm,right=0.75cm of cr2] {\scriptsize CR$(3)$};
\node (cr4)  [circle,draw,minimum height=1.1cm,right=0.75cm of cr3] {\scriptsize CR$(4)$};

\node (zw) [circle,draw,minimum height=1.1cm,below=3cm of cr0] {\scriptsize ZW};
\node (bt1)  [circle,draw,minimum height=1.1cm,right=0.75cm of zw] {\scriptsize BT$(1)$};
\node (bt2)  [circle,draw,minimum height=1.1cm,right=0.75cm of bt1] {\scriptsize BT$(2)$};
\node (bt3)  [circle,draw,minimum height=1.1cm,right=0.75cm of bt2] {\scriptsize BT$(3)$};
\node (bt4)  [circle,draw,minimum height=1.1cm,right=0.75cm of bt3] {\scriptsize BT$(4)$};

\path (bt4) edge  [bend left] node [below]{\tiny $1-\xi(4)$} (bt2);
\path (bt3) edge  [bend left] node [below]{\tiny $1-\xi(3)$} (bt1);
\path (bt2) edge  [bend left] node [below]{\tiny $1-\xi(2)$} (zw);

\path (bt1) edge node [above]{\tiny $1\!-\!\xi(1)$} (zw);
\path (zw) edge node [left, very near start]{\tiny $1$} (cr0);

\path (bt1) edge [bend left=10] node [left, very near start]{\tiny $\xi(1)$} (cr0);
\path (bt2) edge [bend left=25] node [below, very near start]{\tiny $\xi(2)$} (cr0);
\path (bt3) edge [bend right=7] node [right]{\tiny $\xi(3)$} (cr1);
\path (bt4) edge [bend right=15] node [right,near end]{\tiny $\xi(4)$} (cr2);

\path (cr4) edge node [right,near end]{\tiny $1$} (bt4);

\path (cr0) edge [bend right=15] node [right]{\tiny $\zeta_0(2)$} (bt2);
\path (cr0) edge [bend right=10]  node [right]{\tiny $\,\zeta_0(3)$} (bt3);\
\path (cr0) edge [bend right=5]  (bt4);
\node (l1) at(1.6,-0.65) {\tiny $1\!-\!\zeta_0(2)$};
\node (l2) at(1.75,-0.9) {\tiny $-\!\zeta_0(3)$};

\path (cr1) edge [bend right=15] node [right, near start]{\tiny $\zeta_1(2)$} (bt3);
\path (cr1) edge [bend left=5] node [right, very near start]{\tiny $1-\zeta_1(2)$} (bt4);

\path (cr2) edge node [right, near start]{\tiny $1$} (bt4);

\path (cr3) edge [bend left=15] node [right]{\tiny $1$} (bt4);

\end{tikzpicture}
\vspace{-0.5cm}
    \caption{Approximated semi-Markov model of the system with $K=3N$ and $\Psi=4$.}\vspace{-0.4cm}
    \label{fig:markov_model}
\end{figure}

\subsection{Belief Threshold Optimization}

We can create a semi-Markov model of the system, as shown in Fig.~\ref{fig:markov_model}, by applying some simplifications: firstly, we consider nodes with the same activation probability $\lambda$. Setting a threshold $F$ on the probability of being the highest node then corresponds to setting a maximum number $K=\frac{\log(F)}{\log(1-\lambda)}$ of possible slots in which the nodes transmit. Secondly, we consider some approximations in the outcomes of the \gls{bt} phase, which we will discuss below.

The \gls{zw} state always leads to a collision, i.e., to a \gls{cr} phase, but the state of the model also keeps track of the highest $\psi^*$ (which is always $0$ for the \gls{zw} phase). Correspondingly, each sequence of \gls{cr} and \gls{ce} phases ends with a transition to the \gls{bt} phase, but $\bm{\psi}$ depends on the duration of the sequence, which we have analyzed above. 
During the \gls{bt} phase, we simplify the model by considering the case in which a single collision resolution phase led to the current state, i.e., by discarding secondary collisions that happen while in the \gls{bt} phase. Given the maximum possible \gls{aoii} $\psi$, we can obtain the conditioned \gls{pmf} of the number of colliders by applying Bayes' theorem:
\begin{equation}
    p_C(C|\psi)=\frac{p_C(c|\text{ZW})p_{\tau_c}(\psi^*)}{p_{\tau}(\psi)},
\end{equation}
where $p_{\tau}(\psi)$ is the \gls{pmf} corresponding to the \gls{cdf} in~\eqref{eq:cr_duration}.

We then consider a pessimistic and an optimistic model. The pessimistic model considers $L(\psi)=N$, i.e., all nodes are considered as possible colliders, independently of their $\psi_{n,t}$. This is a pessimistic estimate, as some nodes might have a lower $\psi_{n,t}$ such that it is common knowledge that they cannot be part of the collision set. On the other hand, the optimistic model subtracts the expected number of colliders from the set of active nodes, considering that they have a much lower \gls{aoi} and, as such, will not transmit. This model is optimistic, as it considers a single collision resolution phase, while the previous dynamics might be more complex and lead to a larger number of potential colliders. The number of active nodes in the optimistic model is $L(\psi)=N-\E{C|\psi}$.
Each sensor transmits with probability $\alpha=1-(1-\lambda)^{\frac{K}{L(\psi)}}$, so the collision probability is
\begin{equation}
\xi(\psi)=1-(1-\lambda)^K-(1-\varepsilon)\Bin\left(1;L(\psi),\alpha\right).
\end{equation}
In the \gls{zw} phase, we have $K=1$. In the \gls{bt} phase, we typically have less than $N$ active nodes, but we need to set $K>N$, as $\psi_{n,t}$ decreases by $\left\lfloor\frac{K}{L(\psi)}\right\rfloor-1$ for each \gls{bt} step, including those whose outcome is a collision. We can also adjust the transmission probability vector $\mb{p}$ of a \gls{cr} cycle following a collision in a \gls{bt} slot, using $1-(1-\lambda)^{\frac{K}{L(\psi)}}$ as an activation probability and finding the solution from Theorem~\ref{th:collision_opt}.

In order to maintain a finite state space $\mc{S}$, we need to set a maximum \gls{aoii} $\Psi$, so that $|\mc{S}|=2\Psi+1$. We can reduce the approximation error as much as possible by considering a large value that will almost never be reached in practice. This analysis can also be used to ascertain the stability of the system: if the steady-state probability of state $\text{CR}(\Psi)$ does not decrease as $\Psi$ increases, the system is unstable.
We can then give the elements of the transition matrix $\mb{M}$ of our model, considering the transitions toward state \gls{zw}:
\begin{equation}
M_{s,\text{ZW}}=
(1-\xi(\psi))\delta(s,\text{BT}(\psi))\mathbbm{1}(K-\psi L(\psi)).
\end{equation}
As $\psi$ is reduced by $\left\lfloor{K}{L(\psi)}\right\rfloor-1$ steps whenever a collision is avoided in the \gls{bt} phase, only \gls{bt} states with a low value of $\psi$ return directly to \gls{zw}. We can compute the transition probabilities to \gls{cr} states as
\begin{equation}
M_{s,\text{CR}(\psi)}{=}
\begin{cases}
1, &s\!=\!\text{ZW},\psi\!=\!0;\\
\xi(\psi'),&s\!=\!\text{BT}(\psi'), \psi'\!=\!\left[\psi{+}1{-}\frac{K}{L(\psi')}\right]^{\mathclap{\ +}} .
\end{cases}
\end{equation}
Finally, we compute the probability of transitioning to the \gls{bt} phase, considering that $\psi$ is limited to $\Psi$:
\begin{equation}
M_{s,\text{BT}(\psi)}=
\begin{cases}
\zeta_{\psi'}(\psi-\psi'), &s\!=\!\text{CR}(\psi');\\
1-\xi(\psi'),&s\!=\!\text{BT}\left(\psi\!+\!1\!-\!\frac{K}{L(\psi')}\right);\\
\ \ \ \sum\limits_{\mathclap{\ell=\Psi-\psi'}}^{\infty}\,\zeta_{\psi'}(\ell),&s\!=\!\text{CR}(\psi'), \psi=\Psi;
\end{cases}
\end{equation}
where $\zeta_{\psi'}(\ell)$ is the \gls{pmf} corresponding to the \gls{cdf} given in~\eqref{eq:cr_duration}, computed using the optimal transmission probability vector $\mb{p}^*(\psi')$. However, as the system is not a Markov chain, but a discrete-time semi-Markov model, we have $T_{\text{ZW},\text{CR}(0)}=\Geo(\xi(0))$, $T(\text{BT}(\psi),s')=1$, and $T(\text{CR}(\psi),\text{BT}(\psi'))=\psi'-\psi$. We also consider a pessimistic approximation: if the collision resolution process leads to state $\text{BT}(\Psi)$, the time in the \gls{cr} state will be $\Psi$, which should be set to a higher value than the time that is reasonably required to resolve a collision.
We can easily obtain the steady-state probability distribution $\bm{\alpha}$
as the solution to the equation $\bm{\alpha}(\mathbf{P}-\mathbf{I})=0$, normalized so that $||\bm{\alpha}||_1=1$. This corresponds to the left eigenvector of $\mathbf{M}$ with eigenvalue 1. The steady-state distribution $\bm{\pi}$ is obtained by weighting $\bm{\alpha}$ by the average sojourn times $\E{T(s,s')}$:
\begin{equation}
 \pi(s)=\frac{\sum_{s'\in\mc{S}}\alpha(s)M(s,s')\E{T(s,s')}}{\sum_{s^*,s^{**}\in\mc{S}} \alpha(s^*)M(s^*,s^{**})\E{T(s^*,s^{**})}},\quad\forall s\in\mc{S}.
\end{equation}
We can then use $\pi(\text{ZW})$ as a proxy for our desired performance and find $K^*=\argmax_{K\in\mathbb{N}\setminus\{0,1\}}\pi(\text{ZW})$.
Alternatively, we can sum the steady-state probabilities of states that do not violate the \gls{aoii} requirement.

\subsection{Dealing with Imperfect Feedback}\label{sec:delta-imperfect}

Theorem~\ref{th:full_knowledge} requires all nodes to be able to perfectly distinguish between ACKs, NACKs, and silent slots. This condition is met by the ideal and noisy feedback models, as the only confusion in the latter is over the identity of the node receiving the ACK. As we will see in the following, this has a negligible effect on performance, unless the number of nodes in the system is very small.

To compute $\bm{\psi}_t$ and synchronize phase transitions, all nodes need to receive an ACK or NACK after each communication slot. In the \gls{zw}, \gls{cr}, and \gls{ce} phases, this issue can be mitigated by adding only $2$ bits to ACK and NACK packets, representing the current phase (with $4$ possible values). The gateway knows the outcome of each transmission, as it is the intended receiver. It can then compute the current phase and piggyback it on ACK and NACK packets. This synchronizes the protocol for these three phases where knowing the phase completely determines a node's behavior; unless the same node misses multiple feedback packets, the anomaly will be quickly solved, and the protocol will work as intended. Mitigation is more complex in the \gls{bt} phase: since computing $f_{n,t}(\bm{\Theta}_{n,t},\bm{\psi}_t)$ requires a full knowledge of what happened in the past, nodes may have slightly different beliefs over the possible states of the system, leading to inconsistent decision-making processes. We will consider a scheme that includes $\max(\bm{\psi}_t)$ in the feedback packets during the \gls{bt} phase, while sensors simply remain in the same phase if they do not receive an acknowledgment packet, relying on the next one to synchronize with the others. This heuristic might not be optimal, but we show that it is robust with respect to feedback errors, as adapting the Bayesian reasoning in the proof of Theorem~\ref{th:full_knowledge} to this case, considering missed feedback packets as a possible cause of the outcome of each slot, is rather complex.

Additionally, the behavior of the \gls{delta} protocol after a feedback message has been missed is as follows:
\begin{itemize}
    \item In the \gls{zw} and \gls{bt} phases, the node behaves as if the slot was successful until the next feedback message allows it to synchronize the protocol phase. While this choice is optimistic, it leads nodes to avoid reducing their transmission probability unnecessarily if they have new information;
    \item In the \gls{cr} phase, the node behaves as if the slot failed until the next feedback message allows it to synchronize the protocol phase. In the \gls{delta}+ variant, the belief over the number of colliders is not updated;
    \item In the \gls{ce} phase, the node assumes there was a collision, waiting for the next feedback message, unless the slot was silent, in which case it moves to the \gls{bt} phase. In the \gls{delta}+ variant, the belief over the number of colliders is not updated.
\end{itemize}
Under the deletion channel feedback model, nodes in the \gls{ce} phase always move to the \gls{bt} phase. The rationale for this design choice is that, while the \gls{cr} and \gls{ce} phase involve contention for the channel, and thus minimizing the additional traffic ensures a faster recovery, the other phases of the protocol try to avoid collisions at all costs, and thus increasing the traffic slightly by behaving more aggressively for a short time will not have a significant effect. Additionally, even causing a collision will trigger a NACK, leading most nodes to synchronize their protocol phase.

\section{Simulation Settings and Results}\label{sec:sim}

This section presents the results of the Monte Carlo simulations meant to validate the performance of the \gls{delta} protocol. Each considered setting was tested over a simulation lasting $10^6$ slots. In the following, the maximum offered system load $\rho=||\bm{\lambda}||_1$ will be considered as the main simulation parameter.\footnote{The code for the protocol and the simulations in this paper is available at \url{https://github.com/signetlabdei/delta_medium_access}}

\subsection{DELTA Optimization and Robustness}

\begin{figure}[t!]
    \centering
\subfloat[$\rho=0.2$.\label{fig:K_opt_02}]
{\begin{tikzpicture}
\begin{axis}[%
width=\ffwidth,
height=\ffheight,
legend style={legend cell align=left, fill opacity=0.6, draw opacity=1, text opacity=1, legend columns=1, align=left, draw=white!15!black, font=\scriptsize, at={(0.97, 0.03)}, anchor=south east},
xlabel style={font=\footnotesize\color{white!15!black}},
ylabel style={font=\footnotesize\color{white!15!black}},
xmajorgrids,
ymajorgrids,
xmin=0,
xmax=120,
xlabel style={font=\color{white!15!black}},
xlabel={$K$},
ymin=0,
ymax=1,
ylabel style={font=\color{white!15!black}},
ylabel={$\pi(\text{ZW})$},
axis background/.style={fill=white}
]
\addplot [color=color0, mark=o, mark repeat=6, mark phase=6, mark options={solid}]
  table[row sep=crcr]{%
4	0\\
5	0\\
6	0\\
7	0\\
8	0\\
9	0\\
10	0\\
11	0\\
12	0\\
13	0\\
14	0\\
15	0\\
16	0\\
17	0\\
18	0\\
19	0\\
20	0\\
21	0\\
22	0\\
23	0\\
24	0\\
25	0\\
26	0\\
27	0\\
28	0\\
29	0\\
30	0\\
31	0\\
32	0\\
33	0\\
34	0\\
35	0\\
36	0\\
37	0\\
38	0\\
39	0\\
40	0.874958471377118\\
41	0.874958471377118\\
42	0.874958471377118\\
43	0.874958471377118\\
44	0.874958471377118\\
45	0.874958471377118\\
46	0.874958471377118\\
47	0.874958471377118\\
48	0.874958471377118\\
49	0.874958471377118\\
50	0.874958471377118\\
51	0.874958471377118\\
52	0.874958471377118\\
53	0.874958471377118\\
54	0.874958471377118\\
55	0.874958471377118\\
56	0.874958471377118\\
57	0.874958471377118\\
58	0.874958471377118\\
59	0.874958471377118\\
60	0.898445265739158\\
61	0.898445265739158\\
62	0.898445265739158\\
63	0.898445265739158\\
64	0.898445265739158\\
65	0.898445265739158\\
66	0.898445265739158\\
67	0.898445265739158\\
68	0.898445265739158\\
69	0.898445265739158\\
70	0.898445265739158\\
71	0.898445265739158\\
72	0.898445265739158\\
73	0.898445265739158\\
74	0.898445265739158\\
75	0.898445265739158\\
76	0.898445265739158\\
77	0.898445265739158\\
78	0.898445265739158\\
79	0.898445265739158\\
80	0.90203767304214\\
81	0.90203767304214\\
82	0.90203767304214\\
83	0.90203767304214\\
84	0.90203767304214\\
85	0.90203767304214\\
86	0.90203767304214\\
87	0.90203767304214\\
88	0.90203767304214\\
89	0.90203767304214\\
90	0.90203767304214\\
91	0.90203767304214\\
92	0.90203767304214\\
93	0.90203767304214\\
94	0.90203767304214\\
95	0.90203767304214\\
96	0.90203767304214\\
97	0.90203767304214\\
98	0.90203767304214\\
99	0.90203767304214\\
100	0.899912959925185\\
101	0.899912959925185\\
102	0.899912959925185\\
103	0.899912959925185\\
104	0.899912959925185\\
105	0.899912959925185\\
106	0.899912959925185\\
107	0.899912959925185\\
108	0.899912959925185\\
109	0.899912959925185\\
110	0.899912959925185\\
111	0.899912959925185\\
112	0.899912959925185\\
113	0.899912959925185\\
114	0.899912959925185\\
115	0.899912959925185\\
116	0.899912959925185\\
117	0.899912959925185\\
118	0.899912959925185\\
119	0.899912959925185\\
120	0.894487155555762\\
};
\addlegendentry{Pessimistic}

\addplot [color=color1,mark=triangle, mark repeat=6, mark phase=4, mark options={solid}]
  table[row sep=crcr]{%
4	0\\
5	0\\
6	0\\
7	0\\
8	0\\
9	0\\
10	0\\
11	0\\
12	0\\
13	0\\
14	0\\
15	0\\
16	0\\
17	0\\
18	0\\
19	0\\
20	0\\
21	0\\
22	0\\
23	0\\
24	0\\
25	0\\
26	0\\
27	0\\
28	0\\
29	0\\
30	0\\
31	0\\
32	0\\
33	0\\
34	0\\
35	0\\
36	0\\
37	0\\
38	0\\
39	0\\
40	0.934693961705878\\
41	0.934693961705878\\
42	0.934693962413373\\
43	0.934693962413373\\
44	0.934693962413373\\
45	0.934694043792408\\
46	0.934694043792408\\
47	0.934694043792408\\
48	0.934696714147245\\
49	0.934696714147245\\
50	0.934696714147245\\
51	0.934735649062303\\
52	0.934735649062303\\
53	0.934735649062303\\
54	0.935476909963444\\
55	0.935476909963444\\
56	0.935476910200076\\
57	0.948833375460702\\
58	0.948833375460702\\
59	0.948833375460702\\
60	0.947748721144794\\
61	0.947748721144794\\
62	0.947748721144794\\
63	0.947748721144794\\
64	0.947749873394304\\
65	0.947749873394304\\
66	0.947749873394304\\
67	0.947749873394304\\
68	0.947761626067456\\
69	0.947761626067456\\
70	0.947761626234507\\
71	0.947761626234507\\
72	0.948083088790963\\
73	0.948083088790963\\
74	0.948083088790963\\
75	0.948083103910563\\
76	0.950486196675255\\
77	0.950486196675255\\
78	0.950486196675255\\
79	0.950486196675255\\
80	0.949899751830215\\
81	0.949899751830215\\
82	0.949899751830215\\
83	0.949899751830215\\
84	0.94989975189191\\
85	0.949905944047836\\
86	0.949905944047836\\
87	0.949905944047836\\
88	0.949905944047836\\
89	0.949905944047836\\
90	0.950003524294519\\
91	0.950003524294519\\
92	0.950003524294519\\
93	0.950003524294519\\
94	0.950003524294519\\
95	0.949339719217241\\
96	0.949340127089354\\
97	0.949340127089354\\
98	0.949340127148468\\
99	0.949340127148468\\
100	0.949005387666066\\
101	0.949005387666066\\
102	0.949007601849668\\
103	0.949007601849668\\
104	0.949007601849668\\
105	0.949007606744283\\
106	0.949007606744283\\
107	0.949007606744283\\
108	0.949195306658823\\
109	0.949195306658823\\
110	0.949195306658823\\
111	0.949195306658823\\
112	0.949195551160636\\
113	0.949195551160636\\
114	0.946591724747103\\
115	0.946591724747103\\
116	0.946591724747103\\
117	0.946591724747103\\
118	0.946591724747103\\
119	0.946593272363286\\
120	0.946399759281894\\
};
\addlegendentry{Optimistic}

\addplot [color=color2,mark=x, mark repeat=3, mark phase = 1, mark options={solid}]
  table[row sep=crcr]{%
20	0\\
22	0\\
24	0\\
26	0\\
28	0\\
30	0\\
32	0\\
34	0\\
36	0\\
38	0\\
40	0.8035033\\
42	0.8419319\\
44	0.8422109\\
46	0.8419293\\
48	0.8420076\\
50	0.8429091\\
52	0.8462923\\
54	0.8461386\\
56	0.8586058\\
58	0.8714559\\
60	0.8819324\\
62	0.8829293\\
64	0.8827094\\
66	0.8824939\\
68	0.8823419\\
70	0.8831894\\
72	0.8832415\\
74	0.8848662\\
76	0.8894811\\
78	0.8931677\\
80	0.8939966\\
82	0.8935868\\
84	0.8910703\\
86	0.8934303\\
88	0.8934689\\
90	0.8930037\\
92	0.8935655\\
94	0.896391\\
96	0.8951239\\
98	0.8975885\\
100	0.8969829\\
102	0.8971791\\
104	0.8965001\\
106	0.8967775\\
108	0.8967433\\
110	0.8971802\\
112	0.8985448\\
114	0.8983979\\
116	0.8989545\\
118	0.8994145\\
120	0.899601\\
122	0.8989093\\
124	0.8988894\\
126	0.8978762\\
128	0.8989549\\
130	0.899619\\
132	0.8997941\\
134	0.900026\\
136	0.9001492\\
138	0.9002446\\
140	0.8994802\\
142	0.8996593\\
144	0.8996506\\
146	0.9000456\\
148	0.8996606\\
150	0.8999599\\
152	0.9001806\\
154	0.9006415\\
156	0.9003391\\
158	0.9007622\\
160	0.9000658\\
162	0.8997775\\
164	0.9002333\\
166	0.8977235\\
168	0.9002536\\
170	0.9001346\\
172	0.9001127\\
174	0.9004836\\
176	0.9003434\\
178	0.900498\\
180	0.9003494\\
182	0.8996818\\
184	0.9001253\\
186	0.9002608\\
188	0.9000429\\
190	0.9003052\\
192	0.8997842\\
194	0.9002288\\
196	0.900709\\
198	0.8997784\\
200	0.8994745\\
};
\addlegendentry{Monte Carlo}

\end{axis}

\end{tikzpicture}%
}\hspace{0.3cm}
\subfloat[$\rho=0.5$.\label{fig:K_opt_05}]
{
%
%
\begin{tikzpicture}
\begin{axis}[%
width=\ffwidth,
height=\ffheight,
legend style={legend cell align=left, fill opacity=0.6, draw opacity=1, text opacity=1, legend columns=1, align=left, draw=white!15!black, font=\scriptsize, at={(0.03, 0.97)}, anchor=north west},
xlabel style={font=\footnotesize\color{white!15!black}},
ylabel style={font=\footnotesize\color{white!15!black}},
xmajorgrids,
ymajorgrids,
xmin=0,
xmax=120,
xlabel style={font=\color{white!15!black}},
xlabel={$K$},
ymin=0,
ymax=1,
ylabel style={font=\color{white!15!black}},
ylabel={$\pi(\text{ZW})$},
axis background/.style={fill=white}
]
\addplot [color=color0, mark=o, mark repeat=6, mark phase=6, mark options={solid}]
  table[row sep=crcr]{%
4	0\\
5	0\\
6	0\\
7	0\\
8	0\\
9	0\\
10	0\\
11	0\\
12	0\\
13	0\\
14	0\\
15	0\\
16	0\\
17	0\\
18	0\\
19	0\\
20	0\\
21	0\\
22	0\\
23	0\\
24	0\\
25	0\\
26	0\\
27	0\\
28	0\\
29	0\\
30	0\\
31	0\\
32	0\\
33	0\\
34	0\\
35	0\\
36	0\\
37	0\\
38	0\\
39	0\\
40	0.000508194020477723\\
41	0.000508194020477723\\
42	0.000508194020477723\\
43	0.000508194020477723\\
44	0.000508194020477723\\
45	0.000508194020477723\\
46	0.000508194020477723\\
47	0.000508194020477723\\
48	0.000508194020477723\\
49	0.000508194020477723\\
50	0.000508194020477723\\
51	0.000508194020477723\\
52	0.000508194020477723\\
53	0.000508194020477723\\
54	0.000508194020477723\\
55	0.000508194020477723\\
56	0.000508194020477723\\
57	0.000508194020477723\\
58	0.000508194020477723\\
59	0.000508194020477723\\
60	0.00119796017755196\\
61	0.00119796017755196\\
62	0.00119796017755196\\
63	0.00119796017755196\\
64	0.00119796017755196\\
65	0.00119796017755196\\
66	0.00119796017755196\\
67	0.00119796017755196\\
68	0.00119796017755196\\
69	0.00119796017755196\\
70	0.00119796017755196\\
71	0.00119796017755196\\
72	0.00119796017755196\\
73	0.00119796017755196\\
74	0.00119796017755196\\
75	0.00119796017755196\\
76	0.00119796017755196\\
77	0.00119796017755196\\
78	0.00119796017755196\\
79	0.00119796017755196\\
80	0.000298818969720308\\
81	0.000298818969720308\\
82	0.000298818969720308\\
83	0.000298818969720308\\
84	0.000298818969720308\\
85	0.000298818969720308\\
86	0.000298818969720308\\
87	0.000298818969720308\\
88	0.000298818969720308\\
89	0.000298818969720308\\
90	0.000298818969720308\\
91	0.000298818969720308\\
92	0.000298818969720308\\
93	0.000298818969720308\\
94	0.000298818969720308\\
95	0.000298818969720308\\
96	0.000298818969720308\\
97	0.000298818969720308\\
98	0.000298818969720308\\
99	0.000298818969720308\\
100	0\\
101	0\\
102	0\\
103	0\\
104	0\\
105	0\\
106	0\\
107	0\\
108	0\\
109	0\\
110	0\\
111	0\\
112	0\\
113	0\\
114	0\\
115	0\\
116	0\\
117	0\\
118	0\\
119	0\\
120	0\\
};
\addlegendentry{Pessimistic}

\addplot [color=color1,mark=triangle, mark repeat=6, mark phase=4, mark options={solid}]
  table[row sep=crcr]{%
4	0\\
5	0\\
6	0\\
7	0\\
8	0\\
9	0\\
10	0\\
11	0\\
12	0\\
13	0\\
14	0\\
15	0\\
16	0\\
17	0\\
18	0\\
19	0\\
20	0\\
21	0\\
22	0\\
23	0\\
24	0\\
25	0\\
26	0\\
27	0\\
28	0\\
29	0\\
30	0\\
31	0\\
32	0\\
33	0\\
34	0\\
35	0\\
36	0\\
37	0\\
38	0\\
39	0\\
40	0.198769458426522\\
41	0.198769458426522\\
42	0.19950993089666\\
43	0.19950993089666\\
44	0.19950993089666\\
45	0.203811822015487\\
46	0.203811822015487\\
47	0.203811822015487\\
48	0.210698545497053\\
49	0.210698545497053\\
50	0.210698545497053\\
51	0.217537972691703\\
52	0.217552266523779\\
53	0.217552266523779\\
54	0.238141375435489\\
55	0.238141375435489\\
56	0.236745834843521\\
57	0.257996841995808\\
58	0.257996841995808\\
59	0.257996841995808\\
60	0.230742270246851\\
61	0.230742270246851\\
62	0.230742270246851\\
63	0.230742270246851\\
64	0.216474718773673\\
65	0.216517648488029\\
66	0.216517648488029\\
67	0.216517648488029\\
68	0.205760145249406\\
69	0.205760145249406\\
70	0.198824850827967\\
71	0.198824850827967\\
72	0.170876427159835\\
73	0.170876427159835\\
74	0.170876427159835\\
75	0.147871019618392\\
76	0.152169619080369\\
77	0.152169619080369\\
78	0.152245738690151\\
79	0.152245738690151\\
80	0.127585601309463\\
81	0.127585601309463\\
82	0.127585601309463\\
83	0.127585601309463\\
84	0.114585829708386\\
85	0.104654170307571\\
86	0.104654170307571\\
87	0.104654170307571\\
88	0.104654170307571\\
89	0.104654170307571\\
90	0.084175184113089\\
91	0.0842332604529311\\
92	0.0842332604529311\\
93	0.0842332604529311\\
94	0.0842332604529311\\
95	0.0679914356335332\\
96	0.0564905932800702\\
97	0.0564905932800702\\
98	0.0462287780288139\\
99	0.0462287780288139\\
100	0\\
101	0\\
102	0\\
103	0\\
104	0\\
105	0\\
106	0\\
107	0\\
108	0\\
109	0\\
110	0\\
111	0\\
112	0\\
113	0\\
114	0\\
115	0\\
116	0\\
117	0\\
118	0\\
119	0\\
120	0\\
};
\addlegendentry{Optimistic}

\addplot [color=color2,mark=x, mark repeat=3, mark phase=1,mark options={solid}]
  table[row sep=crcr]{%
20	0\\
22	0\\
24	0\\
26	0\\
28	0\\
30	0\\
32	0\\
34	0\\
36	0\\
38	0\\
40	0.1650911\\
42	0.1781902\\
44	0.1808403\\
46	0.1877668\\
48	0.190637\\
50	0.1956585\\
52	0.2049461\\
54	0.2114422\\
56	0.2230184\\
58	0.2427207\\
60	0.2526176\\
62	0.2509338\\
64	0.2479233\\
66	0.2475395\\
68	0.2458057\\
70	0.243408\\
72	0.235638\\
74	0.235522\\
76	0.2431251\\
78	0.2425665\\
80	0.2371834\\
82	0.2351218\\
84	0.231727\\
86	0.2250837\\
88	0.2192906\\
90	0.2132097\\
92	0.2124428\\
94	0.2145689\\
96	0.2096719\\
98	0.2082592\\
100	0.1997464\\
102	0.1941337\\
104	0.1878307\\
106	0.1841741\\
108	0.1829888\\
110	0.1807779\\
112	0.1795095\\
114	0.1748233\\
116	0.1707237\\
118	0.1660019\\
120	0.157221\\
122	0.1527965\\
124	0.1516479\\
126	0.1494003\\
128	0.1455656\\
130	0.1440872\\
132	0.1410559\\
134	0.1358719\\
136	0.1315578\\
138	0.1289713\\
140	0.1222732\\
142	0.1193882\\
144	0.1179761\\
146	0.114635\\
148	0.1133815\\
150	0.1099364\\
152	0.1066126\\
154	0.1050418\\
156	0.1023104\\
158	0.0990844\\
160	0.0945518\\
162	0.0917325\\
164	0.0896282\\
166	0.0883019\\
168	0.0863783\\
170	0.0835259\\
172	0.0813047\\
174	0.0788818\\
176	0.0775595\\
178	0.0760693\\
180	0.0704525\\
182	0.0682833\\
184	0.0677965\\
186	0.0652141\\
188	0.0632351\\
190	0.0617451\\
192	0.0614877\\
194	0.0599912\\
196	0.0580816\\
198	0.0566059\\
200	0.0523725\\
};
\addlegendentry{Monte Carlo}

\end{axis}

\end{tikzpicture}%
}
    \caption{$\pi(\text{ZW})$ as a function of $K$.}\vspace{-0.4cm}
    \label{fig:K_opt}
\end{figure}

First, we analyze the correctness of the theoretical model and the optimization of the \gls{delta} protocol parameters.

Fig.~\ref{fig:K_opt} shows the value of $\pi(\text{ZW})$, which we can use as a proxy for the stability of the protocol, as a function of the chosen $K$. We used a Monte Carlo simulation to verify the two approximations, and considered a case with a $20$\% offered load and a case with a $50$\% offered load. In both cases, the two semi-Markov models lead to the correct optimization of $K$. However, Fig.~\ref{fig:K_opt_02} shows that the optimistic model tends to be less accurate when the load is low. This is due to the nature of collisions in this case: most of the time, higher values of $\psi$ will be reached due to multiple collisions between few nodes or even wireless channel losses, leading the estimated value of $L(\psi)$ to be too low. In this case, the pessimistic model, which assumes that all nodes have the same $\psi_{n,t}$, is closer to the real results. On the other hand, the opposite is true when $\rho=0.5$, as shown in Fig.~\ref{fig:K_opt_05}: when the offered load is high, multiple collisions may cause large differences in the nodes' $\psi_{n,t}$ values, so that the pessimistic model foresees a very low probability of remaining in the \gls{zw} phase. In this case, even the optimistic model is too conservative when $K$ is high, as collisions will be frequent enough that nodes will have very different values of $\psi_{n,t}$, but it manages to capture the trend up to the optimal value of $K$, and as such, it can provide a good guideline for system optimization. \gls{delta} is stable with respect to both $K$ and $p$, and thus robust to errors in the estimation of $\rho$ and $\varepsilon$. In the following, we will show the performance of \gls{delta} with optimized parameters, as well as a version with a fixed value $K=\frac{5}{2}N$, to prove that fixed general settings can perform well in a variety of scenarios.

We can also consider the robustness of the parameter choice in the \gls{cr} phase: Fig.~\ref{fig:cr_opt} shows the result of the transmission probability optimization for different load values. We can note that, aside from the case with $\rho=0.15$, the difference between the outcomes is less than $0.05$ for all \gls{cr} rounds: this means that even significant errors in the load estimation will still lead nodes to behave in a very similar way, resulting in a good protocol performance even under parameter uncertainty.

\subsection{Benchmark Protocols}

We consider two common centralized scheduling algorithms and three distributed protocols as benchmarks to test the \gls{delta} protocol's performance against them in terms of worst-case \gls{aoii} minimization. Firstly, we consider \emph{\acrfull{rr}}, the simplest possible scheduling algorithm. It entirely avoids collisions and does not require sensors to listen to feedback packets, as long as they maintain synchronization, but may lead sensors to wait for a long time if the network is large, as the average \gls{aoi} is $\frac{N}{2}$ even with an error-free channel \cite{jiang2019timely}. \gls{rr} is also vulnerable to wireless channel losses, as a lost packet needs to wait for a full round before being retransmitted. We also implement a \emph{\acrfull{maf}} strategy, which is commonly adopted in the \gls{aoi} literature, as it can optimize the average age in multi-source systems~\cite{bedewy2021optimal}. In our case, it is equivalent to \gls{rr} if $\varepsilon=0$, and has the same issues in large networks with many sensors, but it can efficiently deal with wireless channel losses by retransmitting the lost packet immediately. However, this requires all sensors to listen to feedback packets, as they need to know when packet losses occur.

\begin{figure}[t!]
    \centering
\begin{tikzpicture}

\pgfplotstableread{
R   l1  l2  l3  l4  l5
1   0.587463378906250	0.529724121093750	0.497009277343750	0.473083496093750	0.453430175781250
2   0.592224121093750	0.534240722656250	0.501525878906250	 0.477844238281250	0.458435058593750
3   0.597229003906250	0.538879394531250	0.506286621093750	0.482727050781250	0.463562011718750
4   0.602600097656250	0.543884277343750	0.511169433593750	0.487854003906250	0.469055175781250
5   0.608337402343750	0.549255371093750	0.516540527343750	0.493225097656250	0.474792480468750
6   0.614562988281250	0.554992675781250	0.522155761718750	0.499084472656250	0.480773925781250
7   0.621276855468750	0.561218261718750	0.528137207031250	0.505187988281250	0.486999511718750
8   0.628601074218750	0.567932128906250	0.534729003906250	0.511779785156250	0.493835449218750
9   0.636535644531250	0.575378417968750	0.541931152343750	0.518859863281250	0.501037597656250
10  0.645324707031250	0.583557128906250	0.549743652343750	0.526672363281250	0.508972167968750
}\loadedtable;

\begin{axis}[%
width=\fwidth,
height=\ffheight,
ybar,
tick align=inside,
bar width=1.5pt,
legend style={legend cell align=left, fill opacity=1, draw opacity=1, text opacity=1, legend columns=5, align=left, draw=white!15!black, font=\tiny, at={(0.5, 0.02)}, anchor=south},
xlabel style={font=\footnotesize\color{white!15!black}},
ylabel style={font=\footnotesize\color{white!15!black}},
tick label style={font=\scriptsize\color{white!15!black}},
xmajorgrids,
ymajorgrids,
xmin=0.5,
xmax=10.5,
xlabel={Round $i$},
ymin=0,
ymax=0.7,
ylabel={$p_i^*$},
axis background/.style={fill=white}
]

    \addplot[style={barcolor0,fill={white!20!barcolor0}}] table[x=R, y=l1] {\loadedtable};
    \addlegendentry{$\rho=0.15$};
    \addplot[style={barcolor1,fill={white!20!barcolor1}}] table[x=R, y=l2] {\loadedtable}; 
    \addlegendentry{$\rho=0.3$};
    \addplot[style={barcolor2,fill={white!20!barcolor2}}] table[x=R, y=l3] {\loadedtable}; 
    \addlegendentry{$\rho=0.45$};
    \addplot[style={barcolor3,fill={white!20!barcolor3}}] table[x=R, y=l4] {\loadedtable}; 
    \addlegendentry{$\rho=0.6$};
    \addplot[style={barcolor4,fill={white!20!barcolor4}}] table[x=R, y=l5] {\loadedtable}; 
    \addlegendentry{$\rho=0.75$};

    \end{axis}
\end{tikzpicture}
\caption{Optimal transmission probability for each consecutive \gls{cr} round for different values of $\rho$, with $N=20$ and $\varepsilon=0.05$.}\vspace{-0.4cm}
\label{fig:cr_opt}
\end{figure}

The three distributed algorithms are a variation on the \gls{zw} policy, with different collision resolution mechanisms. 
Firstly, nodes with information to send under the \emph{Pure \acrfull{zw}} policy immediately do so with a certain probability $p_1$. If their packets are lost, either due to the wireless channel or to a collision, they keep transmitting with the same probability until they receive an ACK and return to the normal state. This corresponds to a classical slotted ALOHA system. We also consider a \emph{\gls{lzw}} scheme with two distinct probabilities. Each node transmits with probability $p_1$ if it has information to send, then switches to probability $p_2$ after a failure until the packet is successfully transmitted. This corresponds to a local back-off mechanism after collisions with $p_2$-persistence. Both \gls{zw} and \gls{lzw} only require sensors to listen to feedback packets after they transmit.

\begin{figure*}[t!]
    \centering
\subfloat[\gls{aoii} violation probability ($\Theta_{\max}=0$).\label{fig:lambda_aoii_0}]
{\input{./fig/lambda_violation_aoii0.tex}}
\subfloat[\gls{aoii} violation probability ($\Theta_{\max}=5$).\label{fig:lambda_aoii_5}]
{\input{./fig/lambda_violation_aoii5.tex}}
    \caption{\gls{aoii} violation probability as a function of $\rho$, $N=20$.}\vspace{-0.6cm}
    \label{fig:lambda_viol}
\end{figure*}

Finally, the \emph{\gls{gzw}} protocol is similar to \gls{lzw}, but the back-off mechanism is implemented by all nodes. After a transmission failure, all nodes switch from $p_1$ to $p_2$. They then go back to $p_1$ after a successful transmission, assuming the collision involved either $1$ or $2$ nodes. This protocol is fairer than \gls{lzw}, which can lead colliding nodes to have a lower priority than other nodes with a lower \gls{aoii}, but requires all nodes to listen to the feedback for every slot.

The values of $p_1$ and $p_2$ for the distributed benchmarks were optimized for each specific scenario by performing a grid search over a Markov representation of the protocols.

\begin{figure*}[t!]
    \centering
\subfloat[\gls{aoii} violation probability ($\Theta_{\max}=0$, $\rho=0.3$).\label{fig:N_aoii_0_rho3}]
{
%
%
%
\begin{tikzpicture}

\begin{axis}[%
width=\fwidth,
height=\fheight,
legend style={legend cell align=left, fill opacity=0.6, draw opacity=1, text opacity=1, legend columns=3, align=left, draw=white!15!black, font=\tiny, at={(0.99, 0.98)}, anchor=north east},
xlabel style={font=\footnotesize\color{white!15!black}},
ylabel style={font=\footnotesize\color{white!15!black}},
tick label style={font=\scriptsize\color{white!15!black}},
xmajorgrids,
ymajorgrids,
xmin=4,
xmax=50,
xlabel={Number of nodes $N$},
ymin=0,
ymax=0.3,
ylabel={Violation prob. $V(0)$},
axis background/.style={fill=white}
]
\addplot [color=color0, mark=o, mark repeat=2, mark phase=2, mark options={solid}]
  table{%
4	0.119351000000000
6	0.129197333333333
8	0.133188875000000
10	0.137290200000000
12	0.138267500000000
14	0.140240857142857
16	0.140850125000000
18	0.141626222222222
20	0.142491550000000
22	0.143052545454546
24	0.143532583333333
26	0.143753961538462
28	0.143622285714286
30	0.144417366666667
32	0.144302156250000
34	0.144766382352941
36	0.144877638888889
38	0.144692394736842
40	0.144849425000000
42	0.144568095238095
44	0.144806431818182
46	0.145541217391304
48	0.145468437500000
50	0.146552080000000
};
\addlegendentry{RR}

\addplot [color=color0,mark=triangle, mark repeat=2, mark phase=1, mark options={solid}]
  table{%
4	0.115508000000000
6	0.124974666666667
8	0.129402000000000
10	0.131928500000000
12	0.133546333333333
14	0.134840785714286
16	0.136367750000000
18	0.136416833333333
20	0.137665250000000
22	0.137784000000000
24	0.138352708333333
26	0.138675923076923
28	0.138246535714286
30	0.138731900000000
32	0.139752125000000
34	0.139798147058824
36	0.139483194444444
38	0.139793184210526
40	0.139791325000000
42	0.140443666666667
44	0.139870886363636
46	0.139793760869565
48	0.140532458333333
50	0.140843140000000
};
\addlegendentry{MAF}

\addplot [color=color1,mark=o, mark repeat=2, mark phase=1, mark options={solid}]
  table{%
4	0.457603500000000
6	0.376242000000000
8	0.320225500000000
10	0.281838500000000
12	0.250147166666667
14	0.225867428571429
16	0.208428562500000
18	0.191317277777778
20	0.179331600000000
22	0.166331545454545
24	0.157434500000000
26	0.148414730769231
28	0.140129750000000
30	0.133881766666667
32	0.127043937500000
34	0.121826911764706
36	0.116613583333333
38	0.111247105263158
40	0.108444225000000
42	0.104467738095238
44	0.0998014318181818
46	0.0961413478260870
48	0.0935507083333333
50	0.0909554800000000
};
\addlegendentry{\gls{zw}}

\addplot [color=color1,mark=triangle, mark repeat=2, mark phase=2, mark options={solid}]
  table{%
4	0.209282000000000
6	0.178712333333333
8	0.153351750000000
10	0.136475900000000
12	0.124264666666667
14	0.114011000000000
16	0.105042187500000
18	0.0963017777777778
20	0.0913361000000000
22	0.0860933636363637
24	0.0813252083333333
26	0.0774603076923077
28	0.0736435357142857
30	0.0699412000000000
32	0.0668799062500000
34	0.0643290882352942
36	0.0619000833333333
38	0.0603424473684211
40	0.0572049750000000
42	0.0562209047619048
44	0.0538794318181818
46	0.0531555434782609
48	0.0532498958333333
50	0.0500201200000000
};
\addlegendentry{LZW}

\addplot [color=color1,mark=x, mark repeat=2, mark phase=2, mark options={solid}]
  table{%
4	0.264580750000000
6	0.218655833333333
8	0.193252750000000
10	0.173855100000000
12	0.160186416666667
14	0.148916714285714
16	0.141803500000000
18	0.132903777777778
20	0.127472950000000
22	0.122174863636364
24	0.116807541666667
26	0.113734192307692
28	0.110684357142857
30	0.107209733333333
32	0.106490468750000
34	0.101395794117647
36	0.100822666666667
38	0.0991076578947369
40	0.0980401500000000
42	0.0959553333333333
44	0.0971393409090909
46	0.0968343695652174
48	0.0968874583333333
50	0.0957546200000000
};
\addlegendentry{GZW}

\addplot [color=color2,mark=o, mark repeat=2, mark phase=1, mark options={solid}]
  table{%
4	0.0720360000000000
6	0.0627731666666667
8	0.0564115000000000
10	0.0489278000000000
12	0.0444981666666666
14	0.0411776428571429
16	0.0387051250000000
18	0.0350128333333334
20	0.0331932500000000
22	0.0315938181818182
24	0.0293766666666667
26	0.0278398846153847
28	0.0264282500000000
30	0.0249050333333334
32	0.0227683125000000
34	0.0217119705882353
36	0.0209384722222222
38	0.0198776315789474
40	0.0192503500000000
42	0.0202563095238095
44	0.0196558636363636
46	0.0185565217391305
48	0.0184709166666667
50	0.0182204000000000
};
\addlegendentry{DELTA (pess.)}

\addplot [color=color2,mark=triangle, mark repeat=2, mark phase=2, mark options={solid}]
  table{%
4	0.0712390000000001
6	0.0615711666666666
8	0.0548067500000000
10	0.0498546000000000
12	0.0439664166666667
14	0.0412227857142857
16	0.0380861250000000
18	0.0349673888888888
20	0.0327608500000000
22	0.0296989090909091
24	0.0282810833333333
26	0.0272149230769231
28	0.0249525000000000
30	0.0241530000000000
32	0.0226657500000000
34	0.0214218529411765
36	0.0209728333333333
38	0.0201265000000000
40	0.0193977000000000
42	0.0188703571428571
44	0.0183669090909091
46	0.0179082173913043
48	0.0185162500000000
50	0.0176428000000000
};
\addlegendentry{DELTA (opt.)}

\addplot [color=color2,mark=x, mark repeat=2, mark phase=1, mark options={solid}]
  table{%
4	0.0708717500000000
6	0.0606186666666667
8	0.0543416250000000
10	0.0493533000000000
12	0.0451657500000000
14	0.0413168571428572
16	0.0384920000000000
18	0.0355095555555556
20	0.0332733500000000
22	0.0311871363636363
24	0.0293068750000000
26	0.0275203846153846
28	0.0262703928571428
30	0.0248324666666667
32	0.0241397500000000
34	0.0223444117647059
36	0.0224278611111111
38	0.0214472105263158
40	0.0204212750000000
42	0.0213667619047619
44	0.0187707954545454
46	0.0182909565217392
48	0.0176150208333333
50	0.0171160800000000
};
\addlegendentry{DELTA (fixed)}

\addplot [color=color3,mark=x, mark repeat=2, mark phase=1, mark options={solid}]
  table{%
4	0.0713990000000000
6	0.0620411666666667
8	0.0547091250000000
10	0.0484824000000000
12	0.0446008333333333
14	0.0412015000000000
16	0.0378736250000000
18	0.0356246111111112
20	0.0328675000000001
22	0.0306980454545455
24	0.0290748333333334
26	0.0280538846153846
28	0.0261492142857143
30	0.0253812333333333
32	0.0239368437500000
34	0.0228534117647059
36	0.0215880277777778
38	0.0208939473684211
40	0.0203534500000000
42	0.0193930000000000
44	0.0188347954545455
46	0.0182466739130435
48	0.0176215000000000
50	0.0170424400000000
};
\addlegendentry{DELTA+ (fixed)}

\end{axis}
\end{tikzpicture}%
}
\subfloat[\gls{aoii} violation probability ($\Theta_{\max}=5$, $\rho=0.3$).\label{fig:N_aoii_5_rho3}]
{
%
%
\begin{tikzpicture}

\begin{axis}[%
width=\fwidth,
height=\fheight,
legend style={legend cell align=left, fill opacity=0.6, draw opacity=1, text opacity=1, legend columns=3, align=left, draw=white!15!black, font=\tiny, at={(0.99, 0.98)}, anchor=north east},
xlabel style={font=\footnotesize\color{white!15!black}},
ylabel style={font=\footnotesize\color{white!15!black}},
tick label style={font=\scriptsize\color{white!15!black}},
xmajorgrids,
ymajorgrids,
xmin=4,
xmax=50,
xlabel={Number of nodes $N$},
ymin=0,
ymax=0.3,
ylabel={Violation prob. $V(5)$},
axis background/.style={fill=white}
]
\addplot [color=color0, mark=o, mark repeat=2, mark phase=2, mark options={solid}]
  table{%
4	0.00331075000000003
6	0.00857416666666666
8	0.0243597500000000
10	0.0409943000000000
12	0.0536397500000000
14	0.0646755000000000
16	0.0730214375000000
18	0.0800910000000000
20	0.0862220000000000
22	0.0912018181818182
24	0.0955648333333333
26	0.0989823846153847
28	0.101868928571429
30	0.105140766666667
32	0.107341562500000
34	0.109778411764706
36	0.111668833333333
38	0.113110868421053
40	0.114791800000000
42	0.115906500000000
44	0.117349045454546
46	0.119128000000000
48	0.120143333333333
50	0.122090800000000
};
\addlegendentry{RR}

\addplot [color=color0, mark=triangle, mark repeat=2, mark phase=1, mark options={solid}]
  table{%
4	5.02500000000294e-05
6	0.00297400000000003
8	0.0190041250000000
10	0.0349904000000000
12	0.0483287500000000
14	0.0590182857142857
16	0.0679090625000000
18	0.0745541111111111
20	0.0808162000000000
22	0.0855925909090909
24	0.0899965833333334
26	0.0936353461538462
28	0.0963157857142857
30	0.0992799333333333
32	0.102393000000000
34	0.104464176470588
36	0.106058694444444
38	0.108008552631579
40	0.109575075000000
42	0.111464452380952
44	0.112261636363636
46	0.113292565217391
48	0.114974125000000
50	0.116240760000000
};
\addlegendentry{MAF}

\addplot [color=color1, mark=o, mark repeat=2, mark phase=1, mark options={solid}]
  table{%
4	0.299248750000000
6	0.252355333333333
8	0.218297750000000
10	0.194982400000000
12	0.174601833333333
14	0.158792142857143
16	0.147793250000000
18	0.136355777777778
20	0.128765100000000
22	0.119852272727273
24	0.114026000000000
26	0.107882769230769
28	0.102136107142857
30	0.0980403000000000
32	0.0932367500000000
34	0.0897284705882353
36	0.0860893055555556
38	0.0822139210526316
40	0.0805896750000000
42	0.0778275714285714
44	0.0742657045454546
46	0.0717171086956522
48	0.0699813958333333
50	0.0682019400000000
};
\addlegendentry{\gls{zw}}

\addplot [color=color1, mark=triangle, mark repeat=2, mark phase=2, mark options={solid}]  table{%
4	0.143019250000000
6	0.125358166666667
8	0.109403000000000
10	0.0984111000000000
12	0.0905750833333333
14	0.0837430714285714
16	0.0777218125000000
18	0.0715376666666667
20	0.0682393000000000
22	0.0646439545454546
24	0.0612825833333334
26	0.0585873461538462
28	0.0559407857142857
30	0.0532749666666666
32	0.0510341562500000
34	0.0492403529411765
36	0.0475753333333333
38	0.0466161842105263
40	0.0441051250000000
42	0.0435790952380952
44	0.0418471818181818
46	0.0414207173913044
48	0.0418955000000000
50	0.0390930800000000
};
\addlegendentry{LZW}

\addplot [color=color1, mark=x, mark repeat=2, mark phase=1, mark options={solid}]
  table{%
4	0.119704500000000
6	0.109507166666667
8	0.104369625000000
10	0.0987577000000000
12	0.0949026666666667
14	0.0912292142857143
16	0.0896410000000000
18	0.0856777222222223
20	0.0839684500000000
22	0.0819282272727273
24	0.0794630000000000
26	0.0787356153846154
28	0.0776901071428572
30	0.0761688333333334
32	0.0768522812500000
34	0.0734821764705882
36	0.0741692500000000
38	0.0737071842105264
40	0.0737158750000000
42	0.0726934047619048
44	0.0745989090909091
46	0.0752081086956522
48	0.0761112083333333
50	0.0756529600000000
};
\addlegendentry{GZW}

\addplot [color=color2, mark=o, mark repeat=2, mark phase=1, mark options={solid}]
  table{%
4	0.00900800000000002
6	0.0112491666666666
8	0.0123087500000000
10	0.0115420000000001
12	0.0114864166666667
14	0.0115198571428572
16	0.0114745625000000
18	0.0104448888888888
20	0.0104361500000000
22	0.0102994545454546
24	0.00963616666666667
26	0.00938888461538456
28	0.00919042857142860
30	0.00871266666666670
32	0.00730703124999998
34	0.00706700000000005
36	0.00700163888888894
38	0.00668168421052628
40	0.00655300000000003
42	0.00657511904761909
44	0.00652784090909087
46	0.00608193478260866
48	0.00624960416666665
50	0.00638733999999996
};
\addlegendentry{DELTA (pess.)}

\addplot [color=color2, mark=triangle, mark repeat=2, mark phase=2, mark options={solid}]
  table{%
4	0.00870824999999997
6	0.0107101666666667
8	0.0115472500000000
10	0.0116567000000000
12	0.0111603333333333
14	0.0115402857142857
16	0.0110030625000001
18	0.0105282222222223
20	0.0101618500000000
22	0.00859745454545458
24	0.00856016666666670
26	0.00844607692307697
28	0.00770642857142856
30	0.00777356666666662
32	0.00733121874999998
34	0.00694991176470583
36	0.00704527777777775
38	0.00673365789473679
40	0.00669402500000005
42	0.00624952380952382
44	0.00586286363636368
46	0.00582960869565219
48	0.00625427083333330
50	0.00590919999999995
};
\addlegendentry{DELTA (opt.)}

\addplot [color=color2,mark=x, mark repeat=2, mark phase=1, mark options={solid}]
  table{%
4	0.00806474999999995
6	0.00911216666666670
8	0.00973587499999995
10	0.0101765000000000
12	0.00987383333333336
14	0.00953157142857142
16	0.00951768750000004
18	0.00900822222222220
20	0.00878800000000002
22	0.00853336363636359
24	0.00812124999999997
26	0.00786215384615385
28	0.00759757142857143
30	0.00730236666666662
32	0.00735928124999996
34	0.00677838235294115
36	0.00721791666666671
38	0.00692026315789474
40	0.00662527499999999
42	0.00800823809523810
44	0.00607950000000002
46	0.00607260869565218
48	0.00594010416666668
50	0.00572112000000002
};
\addlegendentry{DELTA (fixed)}

\addplot [color=color3,mark=x, mark repeat=2, mark phase=1, mark options={solid}]
  table{%
4	0.00753749999999998
6	0.00908033333333336
8	0.00934999999999997
10	0.00944230000000002
12	0.00908708333333330
14	0.00921128571428576
16	0.00899950000000005
18	0.00876883333333334
20	0.00815684999999999
22	0.00796404545454543
24	0.00776229166666664
26	0.00780365384615389
28	0.00720792857142860
30	0.00732940000000004
32	0.00705271875000002
34	0.00680161764705878
36	0.00638822222222224
38	0.00634873684210524
40	0.00634179999999995
42	0.00599657142857146
44	0.00603172727272727
46	0.00592228260869565
48	0.00572966666666663
50	0.00553336000000004
};
\addlegendentry{DELTA+ (fixed)}

\end{axis}
\end{tikzpicture}%
}\\ \vspace{-0.2cm}
\subfloat[\gls{aoii} violation probability ($\Theta_{\max}=0$, $\rho=0.5$).\label{fig:N_aoii_0}]
{
%
%
%
\begin{tikzpicture}

\begin{axis}[%
width=\fwidth,
height=\fheight,
legend style={legend cell align=left, fill opacity=0.6, draw opacity=1, text opacity=1, legend columns=3, align=left, draw=white!15!black, font=\tiny, at={(0.99, 0.02)}, anchor=south east},
xlabel style={font=\footnotesize\color{white!15!black}},
ylabel style={font=\footnotesize\color{white!15!black}},
tick label style={font=\scriptsize\color{white!15!black}},
xmajorgrids,
ymajorgrids,
xmin=4,
xmax=50,
xlabel={Number of nodes $N$},
ymin=0,
ymax=0.3,
ylabel={Violation prob. $V(0)$},
axis background/.style={fill=white}
]
\addplot [color=color0, mark=o, mark repeat=2, mark phase=2, mark options={solid}]
  table{%
4	0.190701250000000
6	0.203746500000000
8	0.210657875000000
10	0.213976800000000
12	0.216394916666667
14	0.219175071428571
16	0.219864125000000
18	0.220616944444444
20	0.221626750000000
22	0.221898727272727
24	0.222961500000000
26	0.223976576923077
28	0.224005142857143
30	0.223555166666667
32	0.224498187500000
34	0.224705764705882
36	0.224833527777778
38	0.224659026315790
40	0.225076700000000
42	0.226358952380952
44	0.225504272727273
46	0.225479326086957
48	0.225313937500000
50	0.226367180000000
};
\addlegendentry{RR}

\addplot [color=color0,mark=triangle, mark repeat=2, mark phase=1, mark options={solid}]
  table{%
4	0.184898500000000
6	0.197965000000000
8	0.204296375000000
10	0.208107300000000
12	0.210349666666667
14	0.211815357142857
16	0.213436062500000
18	0.214840722222222
20	0.215671350000000
22	0.216678636363636
24	0.216921083333333
26	0.217029115384615
28	0.217238535714286
30	0.217883533333333
32	0.218212875000000
34	0.218847794117647
36	0.218410250000000
38	0.218812236842105
40	0.219266400000000
42	0.219114309523810
44	0.219945681818182
46	0.219750608695652
48	0.219578958333333
50	0.220303400000000
};
\addlegendentry{MAF}

\addplot [color=color1,mark=o, mark repeat=2, mark phase=1, mark options={solid}]
  table{%
4	0.645537750000000
6	0.569154000000000
8	0.515569875000000
10	0.474567900000000
12	0.443776250000000
14	0.417960357142857
16	0.395757312500000
18	0.379140000000000
20	0.364227050000000
22	0.352200454545455
24	0.342699416666667
26	0.334389038461538
28	0.328444714285714
30	0.323840100000000
32	0.319069250000000
34	0.317153205882353
36	0.318507305555556
38	0.320179210526316
40	0.324585200000000
42	0.336617238095238
44	0.354529090909091
46	0.380707152173913
48	0.437122187500000
50	0.535286540000000
};
\addlegendentry{\gls{zw}}

\addplot [color=color1,mark=triangle, mark repeat=2, mark phase=2, mark options={solid}]
  table{%
4	0.464417250000000
6	0.427891666666667
8	0.400628875000000
10	0.377796500000000
12	0.358965500000000
14	0.347467500000000
16	0.336687937500000
18	0.327116277777778
20	0.320179250000000
22	0.312071090909091
24	0.308967500000000
26	0.305848307692308
28	0.303228428571429
30	0.300651900000000
32	0.299255312500000
34	0.300358088235294
36	0.300184416666667
38	0.302428394736842
40	0.304479750000000
42	0.309661238095238
44	0.320626500000000
46	0.325481934782609
48	0.337391562500000
50	0.346484360000000
};
\addlegendentry{LZW}

\addplot [color=color1,mark=x, mark repeat=2, mark phase=2, mark options={solid}]
  table{%
4	0.586571000000000
6	0.517825166666667
8	0.473109250000000
10	0.439890900000000
12	0.412388333333333
14	0.394811714285714
16	0.378481062500000
18	0.366881222222222
20	0.355584300000000
22	0.348534045454545
24	0.344054125000000
26	0.339521153846154
28	0.336048857142857
30	0.331859900000000
32	0.333467843750000
34	0.333164000000000
36	0.336676055555556
38	0.344154973684211
40	0.349059150000000
42	0.356920380952381
44	0.369278181818182
46	0.376266152173913
48	0.399225395833333
50	0.418281800000000
};
\addlegendentry{GZW}

\addplot [color=color2,mark=o, mark repeat=2, mark phase=1, mark options={solid}]
  table{%
4	0.203398800000000
6	0.209571500000000
8	0.196015550000000
10	0.193844100000000
12	0.191404416666667
14	0.189502714285714
16	0.187633625000000
18	0.184560777777778
20	0.185025650000000
22	0.184203272727273
24	0.183895916666667
26	0.182051038461538
28	0.179961464285714
30	0.178482733333333
32	0.178107031250000
34	0.176084882352941
36	0.175723833333333
38	0.175893000000000
40	0.173165675000000
42	0.174234166666667
44	0.173051045454545
46	0.173649217391304
48	0.173552458333333
50	0.171429780000000
};
\addlegendentry{DELTA (pess.)}

\addplot [color=color2,mark=triangle, mark repeat=2, mark phase=2, mark options={solid}]
  table{%
4	0.192875750000000
6	0.196648333333333
8	0.190762125000000
10	0.190680800000000
12	0.186614583333333
14	0.183775857142857
16	0.181656375000000
18	0.179955055555556
20	0.176091250000000
22	0.178674000000000
24	0.180336500000000
26	0.172900653846154
28	0.173560785714286
30	0.172962700000000
32	0.172511875000000
34	0.171789205882353
36	0.169948638888889
38	0.173267789473684
40	0.168589475000000
42	0.169628142857143
44	0.169313227272727
46	0.170207760869565
48	0.168300083333333
50	0.167489660000000
};
\addlegendentry{DELTA (opt.)}

\addplot [color=color2,mark=x, mark repeat=2, mark phase=1, mark options={solid}]
  table{%
4	0.194079500000000
6	0.193448166666667
8	0.192411250000000
10	0.186992600000000
12	0.186235166666667
14	0.183232571428571
16	0.179731250000000
18	0.178540833333333
20	0.176998550000000
22	0.174584545454546
24	0.172027041666667
26	0.172143000000000
28	0.170955535714286
30	0.166563266666667
32	0.166524812500000
34	0.166542911764706
36	0.165610111111111
38	0.166015078947368
40	0.164389200000000
42	0.164290785714286
44	0.162750727272727
46	0.161404739130435
48	0.164539125000000
50	0.160463840000000
};
\addlegendentry{DELTA (fixed)}

\addplot [color=color3,mark=x, mark repeat=2, mark phase=1, mark options={solid}]
  table{%
4	0.194481000000000
6	0.192999500000000
8	0.188631250000000
10	0.182997100000000
12	0.181289083333333
14	0.176149428571429
16	0.173672500000000
18	0.173538222222222
20	0.171132400000000
22	0.167761954545455
24	0.164564333333333
26	0.163473307692308
28	0.164281321428571
30	0.162998333333333
32	0.163173812500000
34	0.160156529411765
36	0.160004000000000
38	0.157220789473684
40	0.158087000000000
42	0.155828976190476
44	0.157794931818182
46	0.156876500000000
48	0.154932520833333
50	0.154795160000000
};
\addlegendentry{DELTA+ (fixed)}

\end{axis}
\end{tikzpicture}%
}
\subfloat[\gls{aoii} violation probability ($\Theta_{\max}=5$, $\rho=0.5$).\label{fig:N_aoii_5}]
{
%
%
\begin{tikzpicture}

\begin{axis}[%
width=\fwidth,
height=\fheight,
legend style={legend cell align=left, fill opacity=0.6, draw opacity=1, text opacity=1, legend columns=3, align=left, draw=white!15!black, font=\tiny, at={(0.02, 0.5)}, anchor=south west},
xlabel style={font=\footnotesize\color{white!15!black}},
ylabel style={font=\footnotesize\color{white!15!black}},
tick label style={font=\scriptsize\color{white!15!black}},
xmajorgrids,
ymajorgrids,
xmin=4,
xmax=50,
xlabel={Number of nodes $N$},
ymin=0,
ymax=0.3,
ylabel={Violation prob. $V(5)$},
axis background/.style={fill=white}
]
\addplot [color=color0, mark=o, mark repeat=2, mark phase=2, mark options={solid}]
  table{%
4	0.00551800000000002
6	0.0136351666666666
8	0.0393416250000001
10	0.0647441000000000
12	0.0855175833333334
14	0.102491214285714
16	0.115580000000000
18	0.126268666666667
20	0.135439550000000
22	0.142791363636364
24	0.149719875000000
26	0.155754423076923
28	0.160260857142857
30	0.163858333333333
32	0.168079406250000
34	0.171489882352941
36	0.174414166666667
38	0.176791710526316
40	0.179434550000000
42	0.182667023809524
44	0.183759181818182
46	0.185469717391304
48	0.186956958333333
50	0.189382980000000
};
\addlegendentry{RR}

\addplot [color=color0, mark=triangle, mark repeat=2, mark phase=1, mark options={solid}]
  table{%
4	7.97500000000451e-05
6	0.00484816666666665
8	0.0311195000000000
10	0.0569760000000000
12	0.0779920833333333
14	0.0946152142857143
16	0.108178125000000
18	0.119361722222222
20	0.128649000000000
22	0.136587272727273
24	0.142836916666667
26	0.148251038461538
28	0.153025928571429
30	0.157524866666667
32	0.161315656250000
34	0.165026735294118
36	0.167523583333333
38	0.170358526315790
40	0.173130350000000
42	0.175105309523810
44	0.177710500000000
46	0.179309673913044
48	0.180867645833333
50	0.182932660000000
};
\addlegendentry{MAF}

\addplot [color=color1, mark=o, mark repeat=2, mark phase=1, mark options={solid}]
  table{%
4	0.463429250000000
6	0.419266333333333
8	0.387596000000000
10	0.362543300000000
12	0.344092083333333
14	0.327797857142857
16	0.313729625000000
18	0.303819111111111
20	0.294343400000000
22	0.287299954545455
24	0.282007333333333
26	0.277323923076923
28	0.274877428571429
30	0.273184733333333
32	0.271140250000000
34	0.271736147058824
36	0.275564194444444
38	0.279304868421053
40	0.285917500000000
42	0.300209928571429
44	0.320520977272727
46	0.349324934782609
48	0.409564270833333
50	0.513153740000000
};
\addlegendentry{\gls{zw}}

\addplot [color=color1, mark=triangle, mark repeat=2, mark phase=2, mark options={solid}]  table{%
4	0.373332750000000
6	0.350419666666667
8	0.332743875000000
10	0.316793700000000
12	0.303550583333333
14	0.296338357142857
16	0.288988562500000
18	0.282523500000000
20	0.278176650000000
22	0.272410045454546
24	0.271250875000000
26	0.269811307692308
28	0.268712928571429
30	0.267634000000000
32	0.267527468750000
34	0.269814411764706
36	0.270742305555556
38	0.274060421052632
40	0.277134325000000
42	0.283121833333333
44	0.294965954545455
46	0.300723608695652
48	0.313456270833333
50	0.323330680000000
};
\addlegendentry{LZW}

\addplot [color=color1, mark=x, mark repeat=2, mark phase=1, mark options={solid}]
  table{%
4	0.387541500000000
6	0.358257166666667
8	0.339171125000000
10	0.324204900000000
12	0.310496000000000
14	0.303482857142857
16	0.295664375000000
18	0.291035333333333
20	0.285753450000000
22	0.283905136363636
24	0.283766916666667
26	0.283112961538462
28	0.283140535714286
30	0.281957833333333
32	0.286487062500000
34	0.288793941176471
36	0.294736638888889
38	0.304357526315790
40	0.311456600000000
42	0.321363261904762
44	0.335775136363636
46	0.344523934782609
48	0.369588166666667
50	0.390648020000000
};
\addlegendentry{GZW}

\addplot [color=color2, mark=o, mark repeat=2, mark phase=1, mark options={solid}]
  table{%
4	0.0403300500000000
6	0.0652965000000000
8	0.0681539250000000
10	0.0789458000000000
12	0.0871492500000000
14	0.0939721428571428
16	0.0994718750000000
18	0.103153888888889
20	0.108834450000000
22	0.112569000000000
24	0.116487333333333
26	0.118674923076923
28	0.119827964285714
30	0.121539400000000
32	0.123856281250000
34	0.124296470588235
36	0.126339527777778
38	0.128408894736842
40	0.127827875000000
42	0.130618571428571
44	0.131038568181818
46	0.133000739130435
48	0.134203729166667
50	0.133582020000000
};
\addlegendentry{DELTA (pess.)}

\addplot [color=color2, mark=triangle, mark repeat=2, mark phase=2, mark options={solid}]
  table{%
4	0.0306955000000000
6	0.0531838333333333
8	0.0639153750000000
10	0.0754188000000000
12	0.0811605833333333
14	0.0864992857142857
16	0.0922421250000000
18	0.0971424444444444
20	0.0993579500000000
22	0.106597909090909
24	0.112929000000000
26	0.109397076923077
28	0.113370714285714
30	0.115943100000000
32	0.118075718750000
34	0.119824147058824
36	0.120575055555556
38	0.125602315789474
40	0.123145025000000
42	0.125756523809524
44	0.127197931818182
46	0.129602260869565
48	0.129099125000000
50	0.129504020000000
};
\addlegendentry{DELTA (opt.)}

\addplot [color=color2,mark=x, mark repeat=2, mark phase=1, mark options={solid}]
  table{%
4	0.0315600000000000
6	0.0490858333333334
8	0.0628188750000001
10	0.0706801000000000
12	0.0797121666666667
14	0.0854080714285714
16	0.0894451250000000
18	0.0945679444444445
20	0.0985608000000000
22	0.101486545454546
24	0.102954375000000
26	0.106870653846154
28	0.108899678571429
30	0.107855800000000
32	0.110930375000000
34	0.113318911764706
36	0.114780305555556
38	0.117213394736842
40	0.117506100000000
42	0.119358142857143
44	0.119510590909091
46	0.119867326086957
48	0.124282291666667
50	0.121751460000000
};
\addlegendentry{DELTA (fixed)}

\addplot [color=color3,mark=x, mark repeat=2, mark phase=1, mark options={solid}]
  table{%
4	0.0304535000000000
6	0.0471868333333333
8	0.0590381250000001
10	0.0663083000000000
12	0.0748232500000000
14	0.0785651428571429
16	0.0834978750000001
18	0.0889758888888889
20	0.0922775000000000
22	0.0942825000000001
24	0.0954257916666667
26	0.0982809615384616
28	0.102276750000000
30	0.103952933333333
32	0.106937718750000
34	0.106740235294118
36	0.108967944444444
38	0.108257473684211
40	0.111155275000000
42	0.110674523809524
44	0.114226272727273
46	0.115028521739130
48	0.114437375000000
50	0.115814920000000
};
\addlegendentry{DELTA+ (fixed)}

\end{axis}
\end{tikzpicture}%
}\\
    \caption{\gls{aoii} violation as a function of $N$.}\vspace{-0.4cm}
    \label{fig:N_viol}
\end{figure*}

\subsection{Performance Evaluation: Ideal Feedback}

We consider the performance of the protocols under the ideal feedback model by measuring the \gls{aoii} violation probability $V(\Theta_{\max})$, which corresponds to the fraction of time that the nodes spend with an \gls{aoii} higher than the threshold value $\Theta_{\max}$. We analyzed the performance with $\Theta_{\max}=0$, which requires nodes to immediately report anomalies, and $\Theta_{\max}=5$, which allows for a short delay before the gateway is successfully informed of the anomaly. Unless otherwise stated, we consider a system with $N=20$ nodes, a channel erasure probability $\varepsilon=0.05$.

Fig.~\ref{fig:lambda_viol} shows the violation probability as a function of the offered load $\rho=||\mb{\lambda}||_1$, i.e., the load on the system if all nodes immediately transmit successfully, which is an upper bound on the actual system load. The plot clearly shows that \gls{delta} outperforms the other random access schemes, which tend to approach the same reliability only for very low values of the offered load. On the other hand, both $V(0)$ and $V(5)$ grow approximately linearly with $\rho$ for \gls{maf} scheduling: as expected, centralized scheduling mechanisms can outperform any random access scheme for congested networks, but \gls{delta} manages to outperform \gls{maf} for $\rho<0.55$, which is a significant improvement over the \gls{zw} benchmark, as well as a very intense traffic for anomaly reporting applications. The performance of the optimistic, pessimistic, and fixed (with $K=50$) variants remains almost the same, and a small difference can be seen only for very high loads. Additionally, the \gls{delta}+ variant is slightly better, but the more intelligent collision resolution mechanism only has a limited effect on the final performance of the protocol. On the other hand, the other random access protocols have a much higher sensitivity to parameter changes, and the jumps for small changes in $\rho$ are due to the quantization of $p_1$ and $p_2$, for which the grid search optimization considered a $0.01$ step.

\begin{figure*}[t!]
    \centering
\subfloat[\gls{aoii} violation probability ($\Theta_{\max}=0$).\label{fig:nu_aoii_0}]
{
%
%
\begin{tikzpicture}

\begin{axis}[%
width=\fwidth,
height=\fheight,
legend style={legend cell align=left, fill opacity=0.6, draw opacity=1, text opacity=1, legend columns=4, align=left, draw=white!15!black, font=\tiny, at={(0.99, 0.015)}, anchor=south east},
xlabel style={font=\footnotesize\color{white!15!black}},
ylabel style={font=\footnotesize\color{white!15!black}},
tick label style={font=\scriptsize\color{white!15!black}},
xmajorgrids,
ymajorgrids,
xmin=0,
xmax=0.5,
xlabel={$\nu$},
ymin=0,
ymax=0.3,
ylabel={Violation prob. $V(0)$},
axis background/.style={fill=white}
]
\addplot [color=color0, mark=o, mark repeat=2, mark phase=2, mark options={solid}]
  table{%
0	0.221606650000000
0.0500000000000000	0.221530700000000
0.100000000000000	0.221281350000000
0.150000000000000	0.221641200000000
0.200000000000000	0.221966900000000
0.250000000000000	0.221307700000000
0.300000000000000	0.222021850000000
0.350000000000000	0.221460400000000
0.400000000000000	0.222604450000000
0.450000000000000	0.221669050000000
0.500000000000000	0.221282900000000
};
\addlegendentry{RR}

\addplot [color=color0,mark=triangle, mark repeat=2, mark phase=1, mark options={solid}]
  table{%
0	0.215164800000000
0.0500000000000000	0.215557100000000
0.100000000000000	0.215348000000000
0.150000000000000	0.214865950000000
0.200000000000000	0.215962100000000
0.250000000000000	0.215425900000000
0.300000000000000	0.214970550000000
0.350000000000000	0.215682600000000
0.400000000000000	0.215387400000000
0.450000000000000	0.215233650000000
0.500000000000000	0.214932300000000
};
\addlegendentry{MAF}

\addplot [color=color1,mark=o, mark repeat=2, mark phase=1, mark options={solid}]
  table{%
0	0.391516550000000
0.0500000000000000	0.383438200000000
0.100000000000000	0.381124900000000
0.150000000000000	0.393920250000000
0.200000000000000	0.385411950000000
0.250000000000000	0.397398350000000
0.300000000000000	0.390112800000000
0.350000000000000	0.390574700000000
0.400000000000000	0.384058150000000
0.450000000000000	0.382977700000000
0.500000000000000	0.390512050000000
};
\addlegendentry{\gls{zw}}

\addplot [color=color1,mark=triangle, mark repeat=2, mark phase=0, mark options={solid}]
  table{%
0	0.340901650000000
0.0500000000000000	0.342444000000000
0.100000000000000	0.339998500000000
0.150000000000000	0.341643350000000
0.200000000000000	0.344025900000000
0.250000000000000	0.343729300000000
0.300000000000000	0.343947000000000
0.350000000000000	0.339381300000000
0.400000000000000	0.342025000000000
0.450000000000000	0.338300400000000
0.500000000000000	0.341626450000000
};
\addlegendentry{\gls{lzw}}

\addplot [color=color1,mark=x, mark repeat=2, mark phase=1, mark options={solid}]
  table{%
0	0.392743900000000
0.0500000000000000	0.391607500000000
0.100000000000000	0.390908950000000
0.150000000000000	0.389514950000000
0.200000000000000	0.386921150000000
0.250000000000000	0.389072100000000
0.300000000000000	0.389220100000000
0.350000000000000	0.392159650000000
0.400000000000000	0.387033100000000
0.450000000000000	0.393278700000000
0.500000000000000	0.390082950000000
};
\addlegendentry{\gls{gzw}}

\addplot [color=color2,mark=o, mark repeat=2, mark phase=1, mark options={solid}]
  table{%
0	0.175063950000000
0.0500000000000000	0.174647750000000
0.100000000000000	0.176391200000000
0.150000000000000	0.176206200000000
0.200000000000000	0.174321250000000
0.250000000000000	0.174667450000000
0.300000000000000	0.174306350000000
0.350000000000000	0.175662750000000
0.400000000000000	0.176504800000000
0.450000000000000	0.176930700000000
0.500000000000000	0.176614050000000
};
\addlegendentry{\gls{delta}}

\addplot [color=color3,mark=x, mark repeat=2, mark phase=2, mark options={solid}]
  table{%
0	0.169119250000000
0.0500000000000000	0.170929600000000
0.100000000000000	0.169072200000000
0.150000000000000	0.169528550000000
0.200000000000000	0.168658200000000
0.250000000000000	0.170203900000000
0.300000000000000	0.171225350000000
0.350000000000000	0.170912300000000
0.400000000000000	0.171338150000000
0.450000000000000	0.170889100000000
0.500000000000000	0.172733650000000
};
\addlegendentry{\gls{delta}+}

\end{axis}
\end{tikzpicture}%
}
\subfloat[\gls{aoii} violation probability ($\Theta_{\max}=5$).
\label{fig:nu_aoii_5}]
{
%
%
\begin{tikzpicture}

\begin{axis}[%
width=\fwidth,
height=\fheight,
legend style={legend cell align=left, fill opacity=0.6, draw opacity=1, text opacity=1, legend columns=4, align=left, draw=white!15!black, font=\tiny, at={(0.01, 0.5)}, anchor=south west},
xlabel style={font=\footnotesize\color{white!15!black}},
ylabel style={font=\footnotesize\color{white!15!black}},
tick label style={font=\scriptsize\color{white!15!black}},
xmajorgrids,
ymajorgrids,
xmin=0,
xmax=0.5,
xlabel={$\nu$},
ymin=0,
ymax=0.3,
ylabel={Violation prob. $V(5)$},
axis background/.style={fill=white}
]
\addplot [color=color0, mark=o, mark repeat=2, mark phase=2, mark options={solid}]
  table{%
0	0.135467500000000
0.0500000000000000	0.135389250000000
0.100000000000000	0.135293950000000
0.150000000000000	0.135552950000000
0.200000000000000	0.135777500000000
0.250000000000000	0.135169250000000
0.300000000000000	0.135758200000000
0.350000000000000	0.135524150000000
0.400000000000000	0.136188800000000
0.450000000000000	0.135581250000000
0.500000000000000	0.135226100000000
};
\addlegendentry{RR}

\addplot [color=color0, mark=triangle, mark repeat=2, mark phase=1, mark options={solid}]
table{%
0	0.128279650000000
0.0500000000000000	0.128406650000000
0.100000000000000	0.128437700000000
0.150000000000000	0.128031700000000
0.200000000000000	0.128793650000000
0.250000000000000	0.128391100000000
0.300000000000000	0.128049500000000
0.350000000000000	0.128546150000000
0.400000000000000	0.128345650000000
0.450000000000000	0.128295150000000
0.500000000000000	0.128129900000000
};
\addlegendentry{MAF}

\addplot [color=color1, mark=o, mark repeat=2, mark phase=1, mark options={solid}]
  table{%
0	0.327099100000000
0.0500000000000000	0.318615600000000
0.100000000000000	0.316164700000000
0.150000000000000	0.329532500000000
0.200000000000000	0.320861800000000
0.250000000000000	0.333586700000000
0.300000000000000	0.325756350000000
0.350000000000000	0.326137950000000
0.400000000000000	0.319340950000000
0.450000000000000	0.318019800000000
0.500000000000000	0.326108450000000
};
\addlegendentry{\gls{zw}}

\addplot [color=color1, mark=triangle, mark repeat=2, mark phase=2, mark options={solid}]
table{%
0	0.289188500000000
0.0500000000000000	0.290569400000000
0.100000000000000	0.288152000000000
0.150000000000000	0.289804550000000
0.200000000000000	0.292164550000000
0.250000000000000	0.291950450000000
0.300000000000000	0.292119550000000
0.350000000000000	0.287618300000000
0.400000000000000	0.290073000000000
0.450000000000000	0.286526850000000
0.500000000000000	0.289654850000000
};
\addlegendentry{\gls{lzw}}

\addplot [color=color1, mark=x, mark repeat=2, mark phase=1, mark options={solid}]  table{%
0	0.328170050000000
0.0500000000000000	0.326930050000000
0.100000000000000	0.326428300000000
0.150000000000000	0.324901350000000
0.200000000000000	0.322312850000000
0.250000000000000	0.324500850000000
0.300000000000000	0.324638000000000
0.350000000000000	0.327430950000000
0.400000000000000	0.322279900000000
0.450000000000000	0.328790450000000
0.500000000000000	0.325480550000000
};
\addlegendentry{\gls{gzw}}

\addplot [color=color2, mark=o, mark repeat=2, mark phase=1, mark options={solid}]
table{
0	0.0967404500000000
0.0500000000000000	0.0963831000000001
0.100000000000000	0.0979771000000000
0.150000000000000	0.0978924500000000
0.200000000000000	0.0963755000000000
0.250000000000000	0.0965358500000000
0.300000000000000	0.0964737000000000
0.350000000000000	0.0974806000000000
0.400000000000000	0.0978517500000000
0.450000000000000	0.0983680000000000
0.500000000000000	0.0981507000000000
};
\addlegendentry{\gls{delta}}

\addplot [color=color3,mark=x, mark repeat=2, mark phase=2, mark options={solid}]
  table{%
0	0.0907981500000000
0.0500000000000000	0.0923764500000001
0.100000000000000	0.0907602000000000
0.150000000000000	0.0911202000000000
0.200000000000000	0.0907672000000001
0.250000000000000	0.0917381000000000
0.300000000000000	0.0926326000000000
0.350000000000000	0.0925176500000000
0.400000000000000	0.0925055500000001
0.450000000000000	0.0924347500000000
0.500000000000000	0.0939933000000000
};
\addlegendentry{\gls{delta}+}

\end{axis}
\end{tikzpicture}%
}
    \caption{AoII violation as a function of the activation probability range $\nu$ with $\rho=0.5$, $N=20$.}\vspace{-0.4cm}
    \label{fig:sigma_viol}
\end{figure*}

\begin{figure*}[t!]
    \centering
\subfloat[\gls{aoii} violation probability ($\Theta_{\max}=0$, $\rho=0.3$).\label{fig:noise_aoii_0_rho30}]
{\begin{tikzpicture}
\begin{axis}[%
width=\fwidth,
height=\fheight,
legend style={legend cell align=left, fill opacity=0.6, draw opacity=1, text opacity=1, legend columns=4, align=left, draw=white!15!black, font=\tiny, at={(0.99, 0.98)}, anchor=north east},
xlabel style={font=\footnotesize\color{white!15!black}},
ylabel style={font=\footnotesize\color{white!15!black}},
tick label style={font=\scriptsize\color{white!15!black}},
xmajorgrids,
ymajorgrids,
xmin=0,
xmax=5,
xlabel={$\sigma_f$},
ymin=0,
ymax=0.3,
ylabel={Violation prob. $V(0)$},
axis background/.style={fill=white}
]
\addplot [color=color0, mark=o, mark options={solid}]
  table{%
0	0.142280125000000
0.250000000000000	0.142036550000000
0.500000000000000	0.142672875000000
0.750000000000000	0.142569250000000
1	0.142367675000000
1.25000000000000	0.142111225000000
1.50000000000000	0.141902975000000
1.75000000000000	0.142314675000000
2	0.142006250000000
2.25000000000000	0.142179575000000
2.50000000000000	0.142005400000000
2.75000000000000	0.142105650000000
3	0.142615150000000
3.25000000000000	0.142175650000000
3.50000000000000	0.142195375000000
3.75000000000000	0.142554775000000
4	0.142093775000000
4.25000000000000	0.142405875000000
4.50000000000000	0.142295950000000
4.75000000000000	0.142623075000000
5	0.142232000000000
};
\addlegendentry{RR}

\addplot [color=color0, mark=triangle, mark options={solid}]
table{%
0	0.137264375000000
0.250000000000000	0.149451800000000
0.500000000000000	0.240541800000000
0.750000000000000	0.318608925000000
1	0.368123275000000
1.25000000000000	0.401405325000000
1.50000000000000	0.422515750000000
1.75000000000000	0.440033450000000
2	0.451407750000000
2.25000000000000	0.459988925000000
2.50000000000000	0.469039050000000
2.75000000000000	0.475073025000000
3	0.479555100000000
3.25000000000000	0.485679800000000
3.50000000000000	0.489008700000000
3.75000000000000	0.492158200000000
4	0.495802900000000
4.25000000000000	0.499292575000000
4.50000000000000	0.501642575000000
4.75000000000000	0.503516075000000
5	0.506501775000000
};
\addlegendentry{MAF}

\addplot [color=color1, mark=o, mark options={solid}]
  table{%
0	0.135153000000000
0.250000000000000	0.132989700000000
0.500000000000000	0.138830000000000
0.750000000000000	0.134051600000000
1	0.133074100000000
1.25000000000000	0.134337450000000
1.50000000000000	0.136193100000000
1.75000000000000	0.127956400000000
2	0.134717850000000
2.25000000000000	0.131864600000000
2.50000000000000	0.127405500000000
2.75000000000000	0.133728750000000
3	0.133287950000000
3.25000000000000	0.137803750000000
3.50000000000000	0.126104850000000
3.75000000000000	0.143322300000000
4	0.131891000000000
4.25000000000000	0.128811050000000
4.50000000000000	0.125325800000000
4.75000000000000	0.132270150000000
5	0.129322050000000
};
\addlegendentry{\gls{zw}}

\addplot [color=color1, mark=triangle, mark options={solid}]
table{%
0	0.0786672500000000
0.250000000000000	0.0809302000000000
0.500000000000000	0.0762073499999999
0.750000000000000	0.0827574499999999
1	0.0848634500000001
1.25000000000000	0.0776829000000001
1.50000000000000	0.0828596499999998
1.75000000000000	0.0823966000000000
2	0.0774562000000002
2.25000000000000	0.0827595000000001
2.50000000000000	0.0801516500000000
2.75000000000000	0.0814050000000002
3	0.0822538999999999
3.25000000000000	0.0797592000000000
3.50000000000000	0.0795382500000000
3.75000000000000	0.0786314000000000
4	0.0811949999999999
4.25000000000000	0.0841265000000000
4.50000000000000	0.0819588500000001
4.75000000000000	0.0816301500000000
5	0.0811840000000000
};
\addlegendentry{\gls{lzw}}

\addplot [color=color1, mark=x, mark options={solid}]  table{%
0	0.118167300000000
0.250000000000000	0.125784750000000
0.500000000000000	0.117185000000000
0.750000000000000	0.120758650000000
1	0.121417200000000
1.25000000000000	0.123849500000000
1.50000000000000	0.122551150000000
1.75000000000000	0.120096550000000
2	0.119512650000000
2.25000000000000	0.125050600000000
2.50000000000000	0.123163000000000
2.75000000000000	0.121544350000000
3	0.123929150000000
3.25000000000000	0.119257350000000
3.50000000000000	0.119737650000000
3.75000000000000	0.121463050000000
4	0.123014150000000
4.25000000000000	0.121702350000000
4.50000000000000	0.125322550000000
4.75000000000000	0.123508250000000
5	0.124070850000000
};
\addlegendentry{\gls{gzw}}

\addplot [color=color2, mark=o, mark options={solid}]
table{
0	0.0332667500000000
0.250000000000000	0.0332289500000000
0.500000000000000	0.0330643000000002
0.750000000000000	0.0328801499999999
1	0.0331256999999999
1.25000000000000	0.0326561999999999
1.50000000000000	0.0332133500000000
1.75000000000000	0.0334534500000000
2	0.0337812000000001
2.25000000000000	0.0333297000000000
2.50000000000000	0.0331699500000000
2.75000000000000	0.0331891999999999
3	0.0333636500000000
3.25000000000000	0.0330419500000000
3.50000000000000	0.0332550000000000
3.75000000000000	0.0328950000000000
4	0.0328587499999999
4.25000000000000	0.0327546000000000
4.50000000000000	0.0331713500000002
4.75000000000000	0.0341694500000000
5	0.0337388499999999
};
\addlegendentry{\gls{delta}}

\addplot [color=color3, mark=x, mark options={solid}]
table{
0	0.0330677000000000
0.250000000000000	0.0327975999999999
0.500000000000000	0.0325413000000000
0.750000000000000	0.0329442000000000
1	0.0334802000000001
1.25000000000000	0.0331277999999998
1.50000000000000	0.0327028000000000
1.75000000000000	0.0330049500000000
2	0.0323293500000000
2.25000000000000	0.0333163500000000
2.50000000000000	0.0331854500000000
2.75000000000000	0.0334387999999999
3	0.0328727999999999
3.25000000000000	0.0337907000000000
3.50000000000000	0.0334893000000001
3.75000000000000	0.0328949999999999
4	0.0332137500000000
4.25000000000000	0.0333944500000001
4.50000000000000	0.0332385500000000
4.75000000000000	0.0327190999999999
5	0.0332319000000000
};
\addlegendentry{\gls{delta}+}

\end{axis}
\end{tikzpicture}%
}
\subfloat[\gls{aoii} violation probability ($\Theta_{\max}=5$, $\rho=0.3$).\label{fig:noise_aoii_5_rho30}]
{\begin{tikzpicture}
\begin{axis}[%
width=\fwidth,
height=\fheight,
legend style={legend cell align=left, fill opacity=0.6, draw opacity=1, text opacity=1, legend columns=4, align=left, draw=white!15!black, font=\tiny, at={(0.99, 0.98)}, anchor=north east},
xlabel style={font=\footnotesize\color{white!15!black}},
ylabel style={font=\footnotesize\color{white!15!black}},
tick label style={font=\scriptsize\color{white!15!black}},
xmajorgrids,
ymajorgrids,
xmin=0,
xmax=5,
xlabel={$\sigma_f$},
ymin=0,
ymax=0.3,
ylabel={Violation prob. $V(5)$},
axis background/.style={fill=white}
]
\addplot [color=color0, mark=o, mark options={solid}]
  table{%
0	0.0859992250000000
0.250000000000000	0.0858167000000001
0.500000000000000	0.0863007749999999
0.750000000000000	0.0862260250000000
1	0.0860575500000000
1.25000000000000	0.0858666750000000
1.50000000000000	0.0857131500000000
1.75000000000000	0.0860219500000000
2	0.0858264000000000
2.25000000000000	0.0859111500000001
2.50000000000000	0.0857872749999999
2.75000000000000	0.0858571750000001
3	0.0862913500000000
3.25000000000000	0.0859247249999999
3.50000000000000	0.0859816750000000
3.75000000000000	0.0861940999999998
4	0.0858640249999999
4.25000000000000	0.0860578000000000
4.50000000000000	0.0859969999999999
4.75000000000000	0.0862468750000001
5	0.0859257500000000
};
\addlegendentry{RR}

\addplot [color=color0, mark=triangle, mark options={solid}]
table{%
0	0.0806211500000000
0.250000000000000	0.0928820750000000
0.500000000000000	0.187270275000000
0.750000000000000	0.269989875000000
1	0.323089725000000
1.25000000000000	0.358660750000000
1.50000000000000	0.381538275000000
1.75000000000000	0.400146825000000
2	0.412588200000000
2.25000000000000	0.421754525000000
2.50000000000000	0.431440275000000
2.75000000000000	0.437954650000000
3	0.442793775000000
3.25000000000000	0.449266125000000
3.50000000000000	0.452977150000000
3.75000000000000	0.456419300000000
4	0.460269675000000
4.25000000000000	0.464084275000000
4.50000000000000	0.466673750000000
4.75000000000000	0.468754175000000
5	0.471971125000000
};
\addlegendentry{MAF}

\addplot [color=color1, mark=o, mark options={solid}]
  table{%
0	0.0919491000000001
0.250000000000000	0.0895718999999998
0.500000000000000	0.0956308000000000
0.750000000000000	0.0908109500000001
1	0.0897802000000001
1.25000000000000	0.0911564499999999
1.50000000000000	0.0929632500000001
1.75000000000000	0.0847040000000001
2	0.0914227999999999
2.25000000000000	0.0885846000000001
2.50000000000000	0.0837208999999999
2.75000000000000	0.0903081499999999
3	0.0900000999999999
3.25000000000000	0.0945190000000000
3.50000000000000	0.0825189500000000
3.75000000000000	0.100616400000000
4	0.0883786999999999
4.25000000000000	0.0854127500000000
4.50000000000000	0.0819003000000000
4.75000000000000	0.0887193500000000
5	0.0858627000000001
};
\addlegendentry{\gls{zw}}

\addplot [color=color1, mark=triangle, mark options={solid}]
table{%
0	0.0548877999999999
0.250000000000000	0.0567574500000000
0.500000000000000	0.0523214000000000
0.750000000000000	0.0586345000000001
1	0.0607996999999999
1.25000000000000	0.0540538500000001
1.50000000000000	0.0588577000000001
1.75000000000000	0.0580735500000000
2	0.0535129000000000
2.25000000000000	0.0585229999999999
2.50000000000000	0.0563274500000001
2.75000000000000	0.0575490500000000
3	0.0580880500000000
3.25000000000000	0.0558272500000000
3.50000000000000	0.0556010499999999
3.75000000000000	0.0547614999999999
4	0.0573384499999999
4.25000000000000	0.0598538000000001
4.50000000000000	0.0578924500000001
4.75000000000000	0.0574990000000000
5	0.0572854999999999
};
\addlegendentry{\gls{lzw}}

\addplot [color=color1, mark=x, mark options={solid}]  table{%
0	0.0839097499999999
0.250000000000000	0.0915783000000000
0.500000000000000	0.0831515500000001
0.750000000000000	0.0867665500000000
1	0.0872690999999999
1.25000000000000	0.0894428500000001
1.50000000000000	0.0881891499999999
1.75000000000000	0.0858745000000000
2	0.0853659000000001
2.25000000000000	0.0907283000000000
2.50000000000000	0.0889993999999998
2.75000000000000	0.0872373500000000
3	0.0895581500000000
3.25000000000000	0.0851009500000001
3.50000000000000	0.0858620500000001
3.75000000000000	0.0872544000000000
4	0.0887289000000000
4.25000000000000	0.0875084500000000
4.50000000000000	0.0909911500000001
4.75000000000000	0.0891515000000001
5	0.0899701500000001
};
\addlegendentry{\gls{gzw}}

\addplot [color=color2, mark=o, mark options={solid}]
table{
0	0.00875334999999999
0.250000000000000	0.00857039999999998
0.500000000000000	0.00849315000000006
0.750000000000000	0.00848910000000003
1	0.00849719999999998
1.25000000000000	0.00823995000000000
1.50000000000000	0.00860035000000003
1.75000000000000	0.00856690000000004
2	0.00888210000000012
2.25000000000000	0.00858884999999998
2.50000000000000	0.00867574999999998
2.75000000000000	0.00854815000000009
3	0.00861300000000009
3.25000000000000	0.00853075000000003
3.50000000000000	0.00876794999999986
3.75000000000000	0.00822185000000009
4	0.00831049999999989
4.25000000000000	0.00835050000000004
4.50000000000000	0.00853039999999994
4.75000000000000	0.00917820000000014
5	0.00881964999999996
};
\addlegendentry{\gls{delta}}

\addplot [color=color3, mark=x, mark options={solid}]
table{
0	0.00836440000000016
0.250000000000000	0.00818764999999999
0.500000000000000	0.00804004999999997
0.750000000000000	0.00832080000000002
1	0.00849905000000006
1.25000000000000	0.00828154999999997
1.50000000000000	0.00813730000000001
1.75000000000000	0.00816465000000011
2	0.00801194999999999
2.25000000000000	0.00853964999999990
2.50000000000000	0.00835700000000006
2.75000000000000	0.00853609999999994
3	0.00827449999999996
3.25000000000000	0.00874205000000017
3.50000000000000	0.00850005000000009
3.75000000000000	0.00828055000000005
4	0.00845505000000002
4.25000000000000	0.00844290000000003
4.50000000000000	0.00840289999999988
4.75000000000000	0.00823000000000007
5	0.00842875000000010
};
\addlegendentry{\gls{delta}+}

\end{axis}
\end{tikzpicture}%
}\\
\subfloat[\gls{aoii} violation probability ($\Theta_{\max}=0$, $\rho=0.5$).\label{fig:noise_aoii_0}]
{\begin{tikzpicture}
\begin{axis}[%
width=\fwidth,
height=\fheight,
legend style={legend cell align=left, fill opacity=0.6, draw opacity=1, text opacity=1, legend columns=4, align=left, draw=white!15!black, font=\tiny, at={(0.99, 0.02)}, anchor=south east},
xlabel style={font=\footnotesize\color{white!15!black}},
ylabel style={font=\footnotesize\color{white!15!black}},
tick label style={font=\scriptsize\color{white!15!black}},
xmajorgrids,
ymajorgrids,
xmin=0,
xmax=5,
xlabel={$\sigma_f$},
ymin=0,
ymax=0.3,
ylabel={Violation prob. $V(0)$},
axis background/.style={fill=white}
]
\addplot [color=color0, mark=o, mark options={solid}]
  table{%
0	0.221655450000000
0.250000000000000	0.221230800000000
0.500000000000000	0.221678850000000
0.750000000000000	0.221771800000000
1	0.221783350000000
1.25000000000000	0.221426450000000
1.50000000000000	0.221580300000000
1.75000000000000	0.221902250000000
2	0.222067200000000
2.25000000000000	0.221499450000000
2.50000000000000	0.221731550000000
2.75000000000000	0.222167550000000
3	0.221464700000000
3.25000000000000	0.221762350000000
3.50000000000000	0.221430100000000
3.75000000000000	0.221734200000000
4	0.222034000000000
4.25000000000000	0.222377500000000
4.50000000000000	0.221963600000000
4.75000000000000	0.221510400000000
5	0.220968800000000
};
\addlegendentry{RR}

\addplot [color=color0, mark=triangle, mark options={solid}]
table{%
0	0.215178275000000
0.250000000000000	0.232722250000000
0.500000000000000	0.357956650000000
0.750000000000000	0.453095650000000
1	0.507975650000000
1.25000000000000	0.542956375000000
1.50000000000000	0.563294800000000
1.75000000000000	0.577736625000000
2	0.588257350000000
2.25000000000000	0.596263075000000
2.50000000000000	0.601667500000000
2.75000000000000	0.607082325000000
3	0.610521100000000
3.25000000000000	0.614291300000000
3.50000000000000	0.617118300000000
3.75000000000000	0.619693175000000
4	0.621837250000000
4.25000000000000	0.623064525000000
4.50000000000000	0.627198200000000
4.75000000000000	0.627874325000000
5	0.630119175000000
};
\addlegendentry{MAF}

\addplot [color=color1, mark=o, mark options={solid}]
  table{%
0	0.388264250000000
0.250000000000000	0.391577200000000
0.500000000000000	0.388264050000000
0.750000000000000	0.394715800000000
1	0.391936200000000
1.25000000000000	0.382734000000000
1.50000000000000	0.389412500000000
1.75000000000000	0.393478450000000
2	0.384943200000000
2.25000000000000	0.388338700000000
2.50000000000000	0.393553950000000
2.75000000000000	0.389241800000000
3	0.389046250000000
3.25000000000000	0.385148550000000
3.50000000000000	0.391821900000000
3.75000000000000	0.389228500000000
4	0.388307150000000
4.25000000000000	0.392640850000000
4.50000000000000	0.388126500000000
4.75000000000000	0.387880900000000
5	0.388150200000000
};
\addlegendentry{\gls{zw}}

\addplot [color=color1, mark=triangle, mark options={solid}]
table{%
0	0.343233400000000
0.250000000000000	0.344454550000000
0.500000000000000	0.345286250000000
0.750000000000000	0.343764950000000
1	0.343794350000000
1.25000000000000	0.346464050000000
1.50000000000000	0.344042450000000
1.75000000000000	0.343073100000000
2	0.344435850000000
2.25000000000000	0.343706600000000
2.50000000000000	0.348319900000000
2.75000000000000	0.342503500000000
3	0.342683550000000
3.25000000000000	0.344850150000000
3.50000000000000	0.343590600000000
3.75000000000000	0.345113800000000
4	0.344726100000000
4.25000000000000	0.342424400000000
4.50000000000000	0.342598750000000
4.75000000000000	0.347811600000000
5	0.343512100000000
};
\addlegendentry{\gls{lzw}}

\addplot [color=color1, mark=x, mark options={solid}]  table{%
0	0.388467050000000
0.250000000000000	0.391563750000000
0.500000000000000	0.390153200000000
0.750000000000000	0.388716050000000
1	0.389410800000000
1.25000000000000	0.393493300000000
1.50000000000000	0.391244400000000
1.75000000000000	0.389429050000000
2	0.389426200000000
2.25000000000000	0.387119100000000
2.50000000000000	0.390578700000000
2.75000000000000	0.389324750000000
3	0.389138350000000
3.25000000000000	0.393147900000000
3.50000000000000	0.389171650000000
3.75000000000000	0.388354950000000
4	0.391191850000000
4.25000000000000	0.388183200000000
4.50000000000000	0.388012700000000
4.75000000000000	0.390175100000000
5	0.387777000000000
};
\addlegendentry{\gls{gzw}}

\addplot [color=color2, mark=o, mark options={solid}]
table{
0	0.170874300000000
0.250000000000000	0.171422050000000
0.500000000000000	0.168525350000000
0.750000000000000	0.169596250000000
1	0.167224050000000
1.25000000000000	0.169046850000000
1.50000000000000	0.168058150000000
1.75000000000000	0.167971700000000
2	0.170163400000000
2.25000000000000	0.168531300000000
2.50000000000000	0.168166600000000
2.75000000000000	0.166298450000000
3	0.168865900000000
3.25000000000000	0.168086400000000
3.50000000000000	0.169783050000000
3.75000000000000	0.167814350000000
4	0.167416350000000
4.25000000000000	0.168334700000000
4.50000000000000	0.169271500000000
4.75000000000000	0.168129300000000
5	0.167983000000000
};
\addlegendentry{\gls{delta}}

\addplot [color=color3, mark=x, mark options={solid}]
table{
0	0.171050750000000
0.250000000000000	0.170927450000000
0.500000000000000	0.169108850000000
0.750000000000000	0.168544600000000
1	0.167483150000000
1.25000000000000	0.168051150000000
1.50000000000000	0.168565000000000
1.75000000000000	0.166433650000000
2	0.169579850000000
2.25000000000000	0.167489150000000
2.50000000000000	0.170267200000000
2.75000000000000	0.169150150000000
3	0.169071600000000
3.25000000000000	0.169948700000000
3.50000000000000	0.169216450000000
3.75000000000000	0.167102350000000
4	0.167120050000000
4.25000000000000	0.167301300000000
4.50000000000000	0.166065850000000
4.75000000000000	0.167722050000000
5	0.168658550000000
};
\addlegendentry{\gls{delta}+}

\end{axis}
\end{tikzpicture}%
}
\subfloat[\gls{aoii} violation probability ($\Theta_{\max}=5$, $\rho=0.5$).\label{fig:noise_aoii_5}]
{\begin{tikzpicture}
\begin{axis}[%
width=\fwidth,
height=\fheight,
legend style={legend cell align=left, fill opacity=0.6, draw opacity=1, text opacity=1, legend columns=4, align=left, draw=white!15!black, font=\tiny, at={(0.99, 0.55)}, anchor=south east},
xlabel style={font=\footnotesize\color{white!15!black}},
ylabel style={font=\footnotesize\color{white!15!black}},
tick label style={font=\scriptsize\color{white!15!black}},
xmajorgrids,
ymajorgrids,
xmin=0,
xmax=5,
xlabel={$\sigma_f$},
ymin=0,
ymax=0.3,
ylabel={Violation prob. $V(5)$},
axis background/.style={fill=white}
]
\addplot [color=color0, mark=o, mark options={solid}]
  table{%
0	0.135543350000000
0.250000000000000	0.135311200000000
0.500000000000000	0.135650550000000
0.750000000000000	0.135595600000000
1	0.135672300000000
1.25000000000000	0.135443200000000
1.50000000000000	0.135564550000000
1.75000000000000	0.135623900000000
2	0.135866100000000
2.25000000000000	0.135391850000000
2.50000000000000	0.135523700000000
2.75000000000000	0.135884650000000
3	0.135292900000000
3.25000000000000	0.135590500000000
3.50000000000000	0.135366600000000
3.75000000000000	0.135569100000000
4	0.135826850000000
4.25000000000000	0.136044950000000
4.50000000000000	0.135821650000000
4.75000000000000	0.135478050000000
5	0.135035750000000
};
\addlegendentry{RR}

\addplot [color=color0, mark=triangle, mark options={solid}]
table{%
0	0.128277725000000
0.250000000000000	0.146718725000000
0.500000000000000	0.282046050000000
0.750000000000000	0.387779900000000
1	0.449404650000000
1.25000000000000	0.488535275000000
1.50000000000000	0.511662900000000
1.75000000000000	0.527939000000000
2	0.539857950000000
2.25000000000000	0.548850400000000
2.50000000000000	0.554999200000000
2.75000000000000	0.561121200000000
3	0.565033650000000
3.25000000000000	0.569269350000000
3.50000000000000	0.572476550000000
3.75000000000000	0.575310300000000
4	0.577748600000000
4.25000000000000	0.579254175000000
4.50000000000000	0.583792300000000
4.75000000000000	0.584739925000000
5	0.587183825000000
};
\addlegendentry{MAF}

\addplot [color=color1, mark=o, mark options={solid}]
  table{%
0	0.323705600000000
0.250000000000000	0.327248750000000
0.500000000000000	0.323629950000000
0.750000000000000	0.330499300000000
1	0.327588200000000
1.25000000000000	0.317907050000000
1.50000000000000	0.324827950000000
1.75000000000000	0.329183550000000
2	0.320233650000000
2.25000000000000	0.323770650000000
2.50000000000000	0.329482400000000
2.75000000000000	0.324652850000000
3	0.324409850000000
3.25000000000000	0.320525950000000
3.50000000000000	0.327499750000000
3.75000000000000	0.324808900000000
4	0.323880750000000
4.25000000000000	0.328418850000000
4.50000000000000	0.323706050000000
4.75000000000000	0.323403800000000
5	0.323685600000000
};
\addlegendentry{\gls{zw}}

\addplot [color=color1, mark=triangle, mark options={solid}]
table{%
0	0.291409600000000
0.250000000000000	0.292556350000000
0.500000000000000	0.293295300000000
0.750000000000000	0.291856900000000
1	0.291918250000000
1.25000000000000	0.294433750000000
1.50000000000000	0.292215300000000
1.75000000000000	0.291035050000000
2	0.292497100000000
2.25000000000000	0.291816900000000
2.50000000000000	0.296308750000000
2.75000000000000	0.290540200000000
3	0.290870600000000
3.25000000000000	0.292952500000000
3.50000000000000	0.291871350000000
3.75000000000000	0.293201950000000
4	0.292453800000000
4.25000000000000	0.290532350000000
4.50000000000000	0.290669950000000
4.75000000000000	0.295986950000000
5	0.291504750000000
};
\addlegendentry{\gls{lzw}}

\addplot [color=color1, mark=x, mark options={solid}]  table{%
0	0.323700400000000
0.250000000000000	0.327073250000000
0.500000000000000	0.325496700000000
0.750000000000000	0.324023750000000
1	0.324817800000000
1.25000000000000	0.329009500000000
1.50000000000000	0.326586300000000
1.75000000000000	0.324747150000000
2	0.324648250000000
2.25000000000000	0.322406900000000
2.50000000000000	0.326059700000000
2.75000000000000	0.324647650000000
3	0.324459600000000
3.25000000000000	0.328573000000000
3.50000000000000	0.324385500000000
3.75000000000000	0.323691800000000
4	0.326628800000000
4.25000000000000	0.323422050000000
4.50000000000000	0.323323900000000
4.75000000000000	0.325406150000000
5	0.323090600000000
};
\addlegendentry{\gls{gzw}}

\addplot [color=color2, mark=o, mark options={solid}]
table{
0	0.0927998000000000
0.250000000000000	0.0933705000000000
0.500000000000000	0.0906912000000000
0.750000000000000	0.0914101500000001
1	0.0892146000000000
1.25000000000000	0.0908098500000000
1.50000000000000	0.0901354999999999
1.75000000000000	0.0898394000000001
2	0.0916763499999999
2.25000000000000	0.0901676999999999
2.50000000000000	0.0901010000000000
2.75000000000000	0.0886346500000002
3	0.0907889500000000
3.25000000000000	0.0898326000000000
3.50000000000000	0.0915313500000000
3.75000000000000	0.0896305000000000
4	0.0893586499999999
4.25000000000000	0.0899355999999999
4.50000000000000	0.0908300500000000
4.75000000000000	0.0899995000000000
5	0.0898269000000000
};
\addlegendentry{\gls{delta}}

\addplot [color=color3, mark=x, mark options={solid}]
table{
0	0.0926308500000002
0.250000000000000	0.0920788000000001
0.500000000000000	0.0906135000000000
0.750000000000000	0.0901225999999999
1	0.0889564500000000
1.25000000000000	0.0895394500000001
1.50000000000000	0.0901872500000002
1.75000000000000	0.0885140499999999
2	0.0908971500000002
2.25000000000000	0.0891409000000001
2.50000000000000	0.0911942000000000
2.75000000000000	0.0903669499999998
3	0.0906362500000001
3.25000000000000	0.0911000499999999
3.50000000000000	0.0904803000000001
3.75000000000000	0.0885821500000000
4	0.0885889000000000
4.25000000000000	0.0891808500000001
4.50000000000000	0.0878797500000000
4.75000000000000	0.0890861499999999
5	0.0901249000000001
};
\addlegendentry{\gls{delta}+}

\end{axis}
\end{tikzpicture}%
}
    \caption{AoII violation as a function of the feedback noise standard deviation $\sigma_f$ with $N=20$.}
    \label{fig:feedback_noi}
\end{figure*}

We can also consider the performance of the schemes as a function of the number of nodes $N$, considering a scenario with a relatively low load ($\rho=0.3$) and one with a high load ($\rho=0.5$). As Fig.~\ref{fig:N_aoii_0_rho3}-\subref*{fig:N_aoii_5_rho3} show, the performance of random access schemes in the low load scenario tends to improve as the number of nodes grows, while scheduled algorithms gradually degrade due to the longer duration between subsequent transmission opportunities for the same node. We note that \gls{delta} significantly outperforms all other schemes, managing to get $V(5)\leq0.01$ for all settings. As for the varying $\lambda$, the fixed variant (with $K=\frac{5}{2}N$) does not lead to any performance degradation, and the \gls{delta}+ variant has a negligible improvement over the basic version of the protocol. This variation is more noticeable in the high load scenario, shown in Fig.~\ref{fig:N_aoii_0}-\subref*{fig:N_aoii_5}, but still relatively small. Even in this scenario, \gls{delta} is remarkably robust to an increased number of nodes, and $V(0)$ improves as the network size grows, although $V(5)$ tends to increase for larger networks. However, \gls{delta} far outstrips other random access protocols and has a significant performance advantage over scheduled schemes for $N>10$.

\begin{figure*}[t!]
    \centering
\subfloat[\gls{aoii} violation probability ($\Theta_{\max}=0$, $\rho=0.3$).\label{fig:feedback_aoii_0_rho30}]
{\begin{tikzpicture}
\begin{axis}[%
width=\fwidth,
height=\fheight,
legend style={legend cell align=left, fill opacity=0.6, draw opacity=1, text opacity=1, legend columns=4, align=left, draw=white!15!black, font=\tiny, at={(0.99, 0.98)}, anchor=north east},
xlabel style={font=\footnotesize\color{white!15!black}},
ylabel style={font=\footnotesize\color{white!15!black}},
tick label style={font=\scriptsize\color{white!15!black}},
xmajorgrids,
ymajorgrids,
xmin=0,
xmax=0.2,
xlabel={$\varepsilon_f$},
xtick={0,0.05,0.1,0.15,0.2},
xticklabels={0,0.05,0.1,0.15,0.2},
ymin=0,
ymax=0.3,
ylabel={Violation prob. $V(0)$},
axis background/.style={fill=white}
]
\addplot [color=color0, mark=o, mark options={solid}]
  table{%
0	0.142148475000000
0.0200000000000000	0.142391525000000
0.0400000000000000	0.142426450000000
0.0600000000000000	0.142142825000000
0.0800000000000000	0.141996775000000
0.100000000000000	0.141970175000000
0.120000000000000	0.142229925000000
0.140000000000000	0.142319250000000
0.160000000000000	0.142061625000000
0.180000000000000	0.142233575000000
0.200000000000000	0.142203325000000
};
\addlegendentry{RR}

\addplot [color=color0, mark=triangle, mark options={solid}]
table{%
0	0.137243750000000
0.0200000000000000	0.139968475000000
0.0400000000000000	0.142984750000000
0.0600000000000000	0.145740525000000
0.0800000000000000	0.149165425000000
0.100000000000000	0.151908075000000
0.120000000000000	0.155270300000000
0.140000000000000	0.159306075000000
0.160000000000000	0.162277525000000
0.180000000000000	0.166449625000000
0.200000000000000	0.169984925000000
};
\addlegendentry{MAF}

\addplot [color=color1, mark=o, mark options={solid}]
  table{%
0	0.134032580000000
0.0200000000000000	0.145160110000000
0.0400000000000000	0.160158610000000
0.0600000000000000	0.185194470000000
0.0800000000000000	0.209910810000000
0.100000000000000	0.247077380000000
0.120000000000000	0.278215860000000
0.140000000000000	0.335982770000000
0.160000000000000	0.392824590000000
0.180000000000000	0.446589000000000
0.200000000000000	0.531661090000000
};
\addlegendentry{\gls{zw}}

\addplot [color=color1, mark=triangle, mark options={solid}]
table{%
0	0.112051640000000
0.0200000000000000	0.115895970000000
0.0400000000000000	0.121691410000000
0.0600000000000000	0.127683900000001
0.0800000000000000	0.134913470000000
0.100000000000000	0.143947320000000
0.120000000000000	0.156972020000000
0.140000000000000	0.166372600000000
0.160000000000000	0.182799070000000
0.180000000000000	0.204382380000000
0.200000000000000	0.230319780000000
};
\addlegendentry{\gls{lzw}}

\addplot [color=color1, mark=x, mark options={solid}]  table{%
0	0.121706020000000
0.0200000000000000	0.131525710000000
0.0400000000000000	0.149457590000000
0.0600000000000000	0.164008760000000
0.0800000000000000	0.182206290000000
0.100000000000000	0.212362510000000
0.120000000000000	0.238242480000000
0.140000000000000	0.267594930000000
0.160000000000000	0.308891390000000
0.180000000000000	0.338019310000000
0.200000000000000	0.383514460000000
};
\addlegendentry{\gls{gzw}}

\addplot [color=color2, mark=o, mark options={solid}]
table{
0	0.0330705599999998
0.0200000000000000	0.0343065400000000
0.0400000000000000	0.0356201099999997
0.0600000000000000	0.0369726299999997
0.0800000000000000	0.0383581700000002
0.100000000000000	0.0400524700000000
0.120000000000000	0.0416332999999999
0.140000000000000	0.0433212000000000
0.160000000000000	0.0451399899999999
0.180000000000000	0.0473952899999998
0.200000000000000	0.0494499399999998
};
\addlegendentry{\gls{delta}}

\addplot [color=color3, mark=x, mark options={solid}]
table{
0	0.0325097500000000
0.0200000000000000	0.0433113000000001
0.0400000000000000	0.0557587500000001
0.0600000000000000	0.0638996999999999
0.0800000000000000	0.0719014000000001
0.100000000000000	0.0861541500000000
0.120000000000000	0.106214400000000
0.140000000000000	0.143018250000000
0.160000000000000	0.142407950000000
0.180000000000000	0.164248300000000
0.200000000000000	0.175673600000000
};
\addlegendentry{\gls{delta}+}

\end{axis}
\end{tikzpicture}%
}
\subfloat[\gls{aoii} violation probability ($\Theta_{\max}=5$, $\rho=0.3$).\label{fig:feedback_aoii_5_rho30}]
{\begin{tikzpicture}
\begin{axis}[%
width=\fwidth,
height=\fheight,
legend style={legend cell align=left, fill opacity=0.6, draw opacity=1, text opacity=1, legend columns=4, align=left, draw=white!15!black, font=\tiny, at={(0.99, 0.98)}, anchor=north east},
xlabel style={font=\footnotesize\color{white!15!black}},
ylabel style={font=\footnotesize\color{white!15!black}},
tick label style={font=\scriptsize\color{white!15!black}},
xmajorgrids,
ymajorgrids,
xmin=0,
xmax=0.2,
xlabel={$\varepsilon_f$},
xtick={0,0.05,0.1,0.15,0.2},
xticklabels={0,0.05,0.1,0.15,0.2},
ymin=0,
ymax=0.3,
ylabel={Violation prob. $V(5)$},
axis background/.style={fill=white}
]
\addplot [color=color0, mark=o, mark options={solid}]
  table{%
0	0.0858716500000000
0.0200000000000000	0.0860957999999999
0.0400000000000000	0.0860830500000001
0.0600000000000000	0.0859462750000000
0.0800000000000000	0.0858147000000000
0.100000000000000	0.0858624250000000
0.120000000000000	0.0860521000000001
0.140000000000000	0.0860383749999998
0.160000000000000	0.0859058250000001
0.180000000000000	0.0859119250000001
0.200000000000000	0.0859408500000001
};
\addlegendentry{RR}

\addplot [color=color0, mark=triangle, mark options={solid}]
table{%
0	0.0805854500000002
0.0200000000000000	0.0833724499999999
0.0400000000000000	0.0863467500000000
0.0600000000000000	0.0891558250000000
0.0800000000000000	0.0924405999999999
0.100000000000000	0.0953866500000002
0.120000000000000	0.0988272499999999
0.140000000000000	0.102713225000000
0.160000000000000	0.105927075000000
0.180000000000000	0.110054550000000
0.200000000000000	0.113821825000000
};
\addlegendentry{MAF}

\addplot [color=color1, mark=o, mark options={solid}]
  table{%
0	0.0907684000000001
0.0200000000000000	0.102117270000000
0.0400000000000000	0.117543030000000
0.0600000000000000	0.143279950000000
0.0800000000000000	0.168659060000000
0.100000000000000	0.207244160000000
0.120000000000000	0.239412540000000
0.140000000000000	0.299533150000000
0.160000000000000	0.358742080000000
0.180000000000000	0.414993190000000
0.200000000000000	0.503867180000000
};
\addlegendentry{\gls{zw}}

\addplot [color=color1, mark=triangle, mark options={solid}]
table{%
0	0.0666574599999998
0.0200000000000000	0.0703892299999998
0.0400000000000000	0.0761340999999999
0.0600000000000000	0.0821056499999998
0.0800000000000000	0.0894054100000000
0.100000000000000	0.0984380300000001
0.120000000000000	0.111588930000000
0.140000000000000	0.121161090000000
0.160000000000000	0.138081650000000
0.180000000000000	0.160206930000000
0.200000000000000	0.186912050000000
};
\addlegendentry{\gls{lzw}}

\addplot [color=color1, mark=x, mark options={solid}]  table{%
0	0.0874730499999999
0.0200000000000000	0.0966591600000004
0.0400000000000000	0.113844780000000
0.0600000000000000	0.127817990000000
0.0800000000000000	0.145564660000000
0.100000000000000	0.175305920000000
0.120000000000000	0.201088290000000
0.140000000000000	0.230600050000000
0.160000000000000	0.272354070000000
0.180000000000000	0.301776850000000
0.200000000000000	0.348304160000000
};
\addlegendentry{\gls{gzw}}

\addplot [color=color2, mark=o, mark options={solid}]
table{
0	0.00856586000000004
0.0200000000000000	0.00935909000000013
0.0400000000000000	0.0101492299999999
0.0600000000000000	0.0110985400000005
0.0800000000000000	0.0121139900000001
0.100000000000000	0.0132724000000000
0.120000000000000	0.0144531900000003
0.140000000000000	0.0156093200000001
0.160000000000000	0.0170213699999999
0.180000000000000	0.0186515400000000
0.200000000000000	0.0202474400000001
};
\addlegendentry{\gls{delta}}

\addplot [color=color3, mark=x, mark options={solid}]
table{
0	0.00795645000000012
0.0200000000000000	0.0184080000000000
0.0400000000000000	0.0305675000000001
0.0600000000000000	0.0386394500000000
0.0800000000000000	0.0465641000000001
0.100000000000000	0.0607702499999999
0.120000000000000	0.0806943500000000
0.140000000000000	0.117865250000000
0.160000000000000	0.104058700000000
0.180000000000000	0.139063200000000
0.200000000000000	0.150109250000000
};
\addlegendentry{\gls{delta}+}

\end{axis}
\end{tikzpicture}%
}\\
\subfloat[\gls{aoii} violation probability ($\Theta_{\max}=0$, $\rho=0.5$).\label{fig:feedback_aoii_0}]
{\begin{tikzpicture}
\begin{axis}[%
width=\fwidth,
height=\fheight,
legend style={legend cell align=left, fill opacity=0.6, draw opacity=1, text opacity=1, legend columns=4, align=left, draw=white!15!black, font=\tiny, at={(0.99, 0.02)}, anchor=south east},
xlabel style={font=\footnotesize\color{white!15!black}},
ylabel style={font=\footnotesize\color{white!15!black}},
tick label style={font=\scriptsize\color{white!15!black}},
xmajorgrids,
ymajorgrids,
xmin=0,
xmax=0.2,
xlabel={$\varepsilon_f$},
xtick={0,0.05,0.1,0.15,0.2},
xticklabels={0,0.05,0.1,0.15,0.2},
ymin=0,
ymax=0.3,
ylabel={Violation prob. $V(0)$},
axis background/.style={fill=white}
]
\addplot [color=color0, mark=o, mark options={solid}]
  table{%
0	0.221406650000000
0.0200000000000000	0.222027250000000
0.0400000000000000	0.221782825000000
0.0600000000000000	0.221072675000000
0.0800000000000000	0.221909725000000
0.100000000000000	0.221294700000000
0.120000000000000	0.221933775000000
0.140000000000000	0.221712775000000
0.160000000000000	0.222008725000000
0.180000000000000	0.221928050000000
0.200000000000000	0.221445925000000
};
\addlegendentry{RR}

\addplot [color=color0, mark=triangle, mark options={solid}]
table{%
0	0.215568275000000
0.0200000000000000	0.219529925000000
0.0400000000000000	0.223602950000000
0.0600000000000000	0.227784025000000
0.0800000000000000	0.232299000000000
0.100000000000000	0.236715050000000
0.120000000000000	0.241935075000000
0.140000000000000	0.246570650000000
0.160000000000000	0.252096525000000
0.180000000000000	0.256723175000000
0.200000000000000	0.262765950000000
};
\addlegendentry{MAF}

\addplot [color=color1, mark=o, mark options={solid}]
  table{%
0	0.364177675000000
0.0200000000000000	0.376580550000000
0.0400000000000000	0.401633675000000
0.0600000000000000	0.420093325000000
0.0800000000000000	0.441570100000000
0.100000000000000	0.458772750000000
0.120000000000000	0.479678350000000
0.140000000000000	0.498155650000000
0.160000000000000	0.518618875000000
0.180000000000000	0.543615700000000
0.200000000000000	0.561590225000000
};
\addlegendentry{\gls{zw}}

\addplot [color=color1, mark=triangle, mark options={solid}]
table{%
0	0.325742300000000
0.0200000000000000	0.334790340000000
0.0400000000000000	0.343923870000000
0.0600000000000000	0.352246590000000
0.0800000000000000	0.362385530000000
0.100000000000000	0.372681290000000
0.120000000000000	0.383119640000000
0.140000000000000	0.394632750000000
0.160000000000000	0.407075510000000
0.180000000000000	0.419592050000000
0.200000000000000	0.433817490000000
};
\addlegendentry{\gls{lzw}}

\addplot [color=color1, mark=x, mark options={solid}]  table{%
0	0.391705260000000
0.0200000000000000	0.402626240000000
0.0400000000000000	0.415814080000000
0.0600000000000000	0.426382840000000
0.0800000000000000	0.438331760000000
0.100000000000000	0.451990310000000
0.120000000000000	0.463845890000000
0.140000000000000	0.475720780000000
0.160000000000000	0.487379470000000
0.180000000000000	0.499282660000000
0.200000000000000	0.512172720000000
};
\addlegendentry{\gls{gzw}}

\addplot [color=color2, mark=o, mark options={solid}]
table{
0	0.170708860000000
0.0200000000000000	0.178466110000000
0.0400000000000000	0.187539040000000
0.0600000000000000	0.196613560000000
0.0800000000000000	0.205329700000000
0.100000000000000	0.216112230000000
0.120000000000000	0.226124900000000
0.140000000000000	0.235927140000000
0.160000000000000	0.248222620000000
0.180000000000000	0.259339650000000
0.200000000000000	0.270168890000000
};
\addlegendentry{\gls{delta}}

\addplot [color=color3, mark=x, mark options={solid}]
table{
0	0.166663050000000
0.0200000000000000	0.216568100000000
0.0400000000000000	0.276414350000000
0.0600000000000000	0.347266700000000
0.0800000000000000	0.408313400000000
0.100000000000000	0.456098750000000
0.120000000000000	0.520626400000000
0.140000000000000	0.569509550000000
0.160000000000000	0.573123100000000
0.180000000000000	0.638804250000000
0.200000000000000	0.661358750000000
};
\addlegendentry{\gls{delta}+}

\end{axis}
\end{tikzpicture}%
}
\subfloat[\gls{aoii} violation probability ($\Theta_{\max}=5$, $\rho=0.5$).\label{fig:feedback_aoii_5}]
{\begin{tikzpicture}
\begin{axis}[%
width=\fwidth,
height=\fheight,
legend style={legend cell align=left, fill opacity=0.6, draw opacity=1, text opacity=1, legend columns=4, align=left, draw=white!15!black, font=\tiny, at={(0.99, 0.02)}, anchor=south east},
xlabel style={font=\footnotesize\color{white!15!black}},
ylabel style={font=\footnotesize\color{white!15!black}},
tick label style={font=\scriptsize\color{white!15!black}},
xmajorgrids,
ymajorgrids,
xmin=0,
xmax=0.2,
xlabel={$\varepsilon_f$},
xtick={0,0.05,0.1,0.15,0.2},
xticklabels={0,0.05,0.1,0.15,0.2},
ymin=0,
ymax=0.3,
ylabel={Violation prob. $V(5)$},
axis background/.style={fill=white}
]
\addplot [color=color0, mark=o, mark options={solid}]
  table{%
0	0.135403500000000
0.0200000000000000	0.135796825000000
0.0400000000000000	0.135571250000000
0.0600000000000000	0.135125550000000
0.0800000000000000	0.135644050000000
0.100000000000000	0.135176925000000
0.120000000000000	0.135711050000000
0.140000000000000	0.135536900000000
0.160000000000000	0.135704050000000
0.180000000000000	0.135730775000000
0.200000000000000	0.135288575000000
};
\addlegendentry{RR}

\addplot [color=color0, mark=triangle, mark options={solid}]
table{%
0	0.128588875000000
0.0200000000000000	0.132678400000000
0.0400000000000000	0.137029500000000
0.0600000000000000	0.141476575000000
0.0800000000000000	0.146125050000000
0.100000000000000	0.150871725000000
0.120000000000000	0.156156475000000
0.140000000000000	0.161272050000000
0.160000000000000	0.166953850000000
0.180000000000000	0.172104500000000
0.200000000000000	0.178435900000000
};
\addlegendentry{MAF}

\addplot [color=color1, mark=o, mark options={solid}]
  table{%
0	0.299370925000000
0.0200000000000000	0.312741850000000
0.0400000000000000	0.339633150000000
0.0600000000000000	0.359356800000000
0.0800000000000000	0.382421225000000
0.100000000000000	0.401290700000000
0.120000000000000	0.423743800000000
0.140000000000000	0.443935350000000
0.160000000000000	0.466206050000000
0.180000000000000	0.493324350000000
0.200000000000000	0.512922375000000
};
\addlegendentry{\gls{zw}}

\addplot [color=color1, mark=triangle, mark options={solid}]
table{%
0	0.254905800000000
0.0200000000000000	0.264434360000000
0.0400000000000000	0.274078430000000
0.0600000000000000	0.283130470000000
0.0800000000000000	0.293907260000000
0.100000000000000	0.305043850000000
0.120000000000000	0.316208200000000
0.140000000000000	0.328570330000000
0.160000000000000	0.342012360000000
0.180000000000000	0.355533120000000
0.200000000000000	0.370871560000000
};
\addlegendentry{\gls{lzw}}

\addplot [color=color1, mark=x, mark options={solid}]  table{%
0	0.327072220000000
0.0200000000000000	0.338706450000000
0.0400000000000000	0.352703120000000
0.0600000000000000	0.364047490000000
0.0800000000000000	0.376806480000000
0.100000000000000	0.391329770000000
0.120000000000000	0.404167010000000
0.140000000000000	0.416918010000000
0.160000000000000	0.429503620000000
0.180000000000000	0.442515350000000
0.200000000000000	0.456409910000000
};
\addlegendentry{\gls{gzw}}

\addplot [color=color2, mark=o, mark options={solid}]
table{
0	0.0926374799999999
0.0200000000000000	0.0998306700000001
0.0400000000000000	0.108446340000000
0.0600000000000000	0.117130440000000
0.0800000000000000	0.125735390000000
0.100000000000000	0.136314390000000
0.120000000000000	0.146076490000000
0.140000000000000	0.156036690000000
0.160000000000000	0.168272160000000
0.180000000000000	0.179577060000000
0.200000000000000	0.190597840000000
};
\addlegendentry{\gls{delta}}

\addplot [color=color3, mark=x, mark options={solid}]
table{
0	0.0883622500000000
0.0200000000000000	0.142337350000000
0.0400000000000000	0.206079350000000
0.0600000000000000	0.282848200000000
0.0800000000000000	0.348958900000000
0.100000000000000	0.400389450000000
0.120000000000000	0.470434100000000
0.140000000000000	0.524115500000000
0.160000000000000	0.527595200000000
0.180000000000000	0.599553900000000
0.200000000000000	0.623758550000000
};
\addlegendentry{\gls{delta}+}

\end{axis}
\end{tikzpicture}%
}
    \caption{AoII violation as a function of the feedback erasure probability $\varepsilon_f$ with $N=20$.}\vspace{-0.4cm}
    \label{fig:feedback_err}
\end{figure*}

Finally, we consider the robustness to errors in the estimated activation rates: we set a load $\rho=0.5$, and randomly sampled $100$ activation probability vectors $\bm{\lambda}\sim\mc{U}\left(\frac{(1-\nu)\rho}{N},\frac{(1+\nu)\rho}{N}\right)$. The input to \gls{delta} was then the average vector, with growing differences among nodes as $\nu$ increased. The resulting \gls{aoii} violation probability is shown in Fig.~\ref{fig:sigma_viol}: all protocols are robust to this type of disruption, and in particular, \gls{delta} and \gls{delta}+ are insensitive to changes in the activation probabilities, as long as the overall load is approximately correct.

\subsection{Performance Evaluation: Imperfect Feedback}

We then evaluate the robustness of the schemes to imperfect feedback, considering the low load ($\rho=0.3$) and high load ($\rho=0.5$) scenarios with $N=20$ and $\varepsilon=0.05$ and following the three imperfect feedback models outlined in Sec.~\ref{sec:sys}.

We first start with the noisy feedback model, in which ACKs, NACKs, and silent slots can always be distinguished, but nodes may erroneously interpret the content of messages: Fig.~\ref{fig:feedback_noi} shows the \gls{aoii} violation probability as a function of the error standard deviation $\sigma_f$. As the figure clearly shows, all protocols except \gls{maf} are almost unaffected. On the other hand, \gls{maf} is strongly affected by this feedback model, as the feedback messages serve as polling requests: if a node mistakenly believes that it has been polled, it will transmit an update, potentially causing a collision. The performance advantage of \gls{delta} and \gls{delta}+ is unaffected by errors on the feedback, even if they are significant (a standard deviation $\sigma_f=5$ out of a total of $20$ nodes). This is reasonable, as errors on the feedback will affect the belief of nodes only slightly (if at all), considering that there are several nodes that have a high maximum \gls{aoii}. As the algorithm is very robust with respect to the choice of the belief threshold, errors on the identity of the nodes will have a limited effect, as most nodes will still move in lockstep.

\begin{figure*}[t!]
    \centering
\subfloat[\gls{aoii} violation probability ($\Theta_{\max}=0$, $\rho=0.5$).\label{fig:deletion_aoii_0}]
{\begin{tikzpicture}
\begin{axis}[%
width=\fwidth,
height=\fheight,
legend style={legend cell align=left, fill opacity=0.6, draw opacity=1, text opacity=1, legend columns=4, align=left, draw=white!15!black, font=\tiny, at={(0.99, 0.02)}, anchor=south east},
xlabel style={font=\footnotesize\color{white!15!black}},
ylabel style={font=\footnotesize\color{white!15!black}},
tick label style={font=\scriptsize\color{white!15!black}},
xmajorgrids,
ymajorgrids,
xmin=0,
xmax=0.2,
xlabel={$\varepsilon_f$},
xtick={0,0.05,0.1,0.15,0.2},
xticklabels={0,0.05,0.1,0.15,0.2},
ymin=0,
ymax=0.3,
ylabel={Violation prob. $V(0)$},
axis background/.style={fill=white}
]
\addplot [color=color0, mark=o, mark options={solid}]
  table{%
0	0.221734500000000
0.0200000000000000	0.221586375000000
0.0400000000000000	0.221954975000000
0.0600000000000000	0.221679825000000
0.0800000000000000	0.221760575000000
0.100000000000000	0.221320950000000
0.120000000000000	0.221737700000000
0.140000000000000	0.221938500000000
0.160000000000000	0.221463475000000
0.180000000000000	0.222037800000000
0.200000000000000	0.221190775000000
};
\addlegendentry{RR}

\addplot [color=color0, mark=triangle, mark options={solid}]
table{%
0	0.215307350000000
0.0200000000000000	0.219275250000000
0.0400000000000000	0.223536050000000
0.0600000000000000	0.227908550000000
0.0800000000000000	0.232125275000000
0.100000000000000	0.237191600000000
0.120000000000000	0.242218800000000
0.140000000000000	0.246742825000000
0.160000000000000	0.251923075000000
0.180000000000000	0.257036250000000
0.200000000000000	0.263036750000000
};
\addlegendentry{MAF}

\addplot [color=color1, mark=o, mark options={solid}]
  table{%
0	0.391486625000000
0.0200000000000000	0.405784950000000
0.0400000000000000	0.426593900000000
0.0600000000000000	0.448525725000000
0.0800000000000000	0.467117725000000
0.100000000000000	0.486686600000000
0.120000000000000	0.510186700000000
0.140000000000000	0.532029225000000
0.160000000000000	0.549883925000000
0.180000000000000	0.568121450000000
0.200000000000000	0.590070050000000
};
\addlegendentry{\gls{zw}}

\addplot [color=color1, mark=triangle, mark options={solid}]
table{%
0	0.325123200000000
0.0200000000000000	0.334718900000000
0.0400000000000000	0.342441625000000
0.0600000000000000	0.350934900000000
0.0800000000000000	0.359528350000000
0.100000000000000	0.370629100000000
0.120000000000000	0.385175100000000
0.140000000000000	0.395889450000000
0.160000000000000	0.405996050000000
0.180000000000000	0.419396900000000
0.200000000000000	0.433276575000000
};
\addlegendentry{\gls{lzw}}

\addplot [color=color1, mark=x, mark options={solid}]  table{%
0	0.391463300000000
0.0200000000000000	0.402365325000000
0.0400000000000000	0.414768350000000
0.0600000000000000	0.428092175000000
0.0800000000000000	0.441061950000000
0.100000000000000	0.452190100000000
0.120000000000000	0.467524050000000
0.140000000000000	0.479332625000000
0.160000000000000	0.493601875000000
0.180000000000000	0.503126150000000
0.200000000000000	0.514927200000000
};
\addlegendentry{\gls{gzw}}

\addplot [color=color2, mark=o, mark options={solid}]
table{
0	0.171295500000000
0.0200000000000000	0.178173700000000
0.0400000000000000	0.189128175000000
0.0600000000000000	0.197387625000000
0.0800000000000000	0.208376550000000
0.100000000000000	0.219944475000000
0.120000000000000	0.235861175000000
0.140000000000000	0.247101850000000
0.160000000000000	0.259600250000000
0.180000000000000	0.272215325000000
0.200000000000000	0.281768950000000
};
\addlegendentry{\gls{delta}}

\addplot [color=color3, mark=x, mark options={solid}]
table{
0	0.167080075000000
0.0200000000000000	0.213502625000000
0.0400000000000000	0.275841925000000
0.0600000000000000	0.334619150000000
0.0800000000000000	0.399835900000000
0.100000000000000	0.446664175000000
0.120000000000000	0.500315800000000
0.140000000000000	0.536452100000000
0.160000000000000	0.569181825000000
0.180000000000000	0.603345950000000
0.200000000000000	0.631205850000000
};
\addlegendentry{\gls{delta}+}

\end{axis}
\end{tikzpicture}%
}
\subfloat[\gls{aoii} violation probability ($\Theta_{\max}=5$, $\rho=0.5$).\label{fig:deletion_aoii_5}]
{\begin{tikzpicture}
\begin{axis}[%
width=\fwidth,
height=\fheight,
legend style={legend cell align=left, fill opacity=0.6, draw opacity=1, text opacity=1, legend columns=4, align=left, draw=white!15!black, font=\tiny, at={(0.99, 0.02)}, anchor=south east},
xlabel style={font=\footnotesize\color{white!15!black}},
ylabel style={font=\footnotesize\color{white!15!black}},
tick label style={font=\scriptsize\color{white!15!black}},
xmajorgrids,
ymajorgrids,
xmin=0,
xmax=0.2,
xlabel={$\omega_f$},
xtick={0,0.05,0.1,0.15,0.2},
xticklabels={0,0.05,0.1,0.15,0.2},
ymin=0,
ymax=0.3,
ylabel={Violation prob. $V(5)$},
axis background/.style={fill=white}
]
\addplot [color=color0, mark=o, mark options={solid}]
  table{%
0	0.135615950000000
0.0200000000000000	0.135408025000000
0.0400000000000000	0.135704550000000
0.0600000000000000	0.135480400000000
0.0800000000000000	0.135631175000000
0.100000000000000	0.135328275000000
0.120000000000000	0.135587675000000
0.140000000000000	0.135712750000000
0.160000000000000	0.135404150000000
0.180000000000000	0.135807950000000
0.200000000000000	0.135186625000000
};
\addlegendentry{RR}

\addplot [color=color0, mark=triangle, mark options={solid}]
table{%
0	0.128319800000000
0.0200000000000000	0.132513400000000
0.0400000000000000	0.136875725000000
0.0600000000000000	0.141533450000000
0.0800000000000000	0.145945950000000
0.100000000000000	0.151229325000000
0.120000000000000	0.156436525000000
0.140000000000000	0.161327300000000
0.160000000000000	0.166807075000000
0.180000000000000	0.172371075000000
0.200000000000000	0.178593250000000
};
\addlegendentry{MAF}

\addplot [color=color1, mark=o, mark options={solid}]
  table{%
0	0.327098725000000
0.0200000000000000	0.342570475000000
0.0400000000000000	0.364935950000000
0.0600000000000000	0.388535375000000
0.0800000000000000	0.408908500000000
0.100000000000000	0.430057450000000
0.120000000000000	0.455597000000000
0.140000000000000	0.479481450000000
0.160000000000000	0.498865725000000
0.180000000000000	0.518976775000000
0.200000000000000	0.542918525000000
};
\addlegendentry{\gls{zw}}

\addplot [color=color1, mark=triangle, mark options={solid}]
table{%
0	0.254143600000000
0.0200000000000000	0.264331000000000
0.0400000000000000	0.272654700000000
0.0600000000000000	0.281781450000000
0.0800000000000000	0.290885875000000
0.100000000000000	0.302911200000000
0.120000000000000	0.318306100000000
0.140000000000000	0.329829825000000
0.160000000000000	0.340720950000000
0.180000000000000	0.355372450000000
0.200000000000000	0.370337175000000
};
\addlegendentry{\gls{lzw}}

\addplot [color=color1, mark=x, mark options={solid}]  table{%
0	0.326799175000000
0.0200000000000000	0.338558525000000
0.0400000000000000	0.351754225000000
0.0600000000000000	0.366103425000000
0.0800000000000000	0.379873850000000
0.100000000000000	0.391937475000000
0.120000000000000	0.408256600000000
0.140000000000000	0.421089175000000
0.160000000000000	0.436556200000000
0.180000000000000	0.446876350000000
0.200000000000000	0.459780225000000
};
\addlegendentry{\gls{gzw}}

\addplot [color=color2, mark=o, mark options={solid}]
table{
0	0.0930442250000001
0.0200000000000000	0.0999047750000001
0.0400000000000000	0.110008750000000
0.0600000000000000	0.118260050000000
0.0800000000000000	0.129061725000000
0.100000000000000	0.140485125000000
0.120000000000000	0.155853975000000
0.140000000000000	0.167574625000000
0.160000000000000	0.180301325000000
0.180000000000000	0.193183075000000
0.200000000000000	0.203180325000000
};
\addlegendentry{\gls{delta}}

\addplot [color=color3, mark=x, mark options={solid}]
table{
0	0.0886611750000000
0.0200000000000000	0.138354300000000
0.0400000000000000	0.204896600000000
0.0600000000000000	0.268490150000000
0.0800000000000000	0.339220075000000
0.100000000000000	0.389700525000000
0.120000000000000	0.447729900000000
0.140000000000000	0.486886900000000
0.160000000000000	0.522687525000000
0.180000000000000	0.559592250000000
0.200000000000000	0.590340250000000
};
\addlegendentry{\gls{delta}+}

\end{axis}
\end{tikzpicture}%
}
    \caption{AoII violation as a function of the feedback deletion probability $\omega_f$ with $N=20$.}
    \label{fig:feedback_del}
\end{figure*}

We then consider the erasure feedback model: Fig.~\ref{fig:feedback_err} shows performance as a function of the erasure probability $\varepsilon_f$. After including the adaptation of feedback messages discussed in Sec.~\ref{sec:delta-imperfect}, the protocol degrades gracefully in the low load scenario shown in Fig.~\ref{fig:feedback_aoii_0_rho30}-\subref*{fig:feedback_aoii_5_rho30}, maintaining a significant advantage over all other schemes. On the other hand, the \gls{delta}+ variant degrades much faster, as its reliance on acknowledgments to optimize the transmission probability in the \gls{cr} phase leads it to make significant mistakes if the feedback is completely missed. Even if we consider the high load scenario, which is already close to \gls{delta}'s saturation point, with collisions becoming a frequent occurrence, the protocol still comes out on top  for $\varepsilon_f\leq0.1$, as shown in Fig.~\ref{fig:feedback_aoii_0}-\subref*{fig:feedback_aoii_5}. On the other hand, the performance of \gls{delta}+ quickly degrades, becoming even worse than other random access schemes.

Finally, Fig.~\ref{fig:feedback_del} shows the performance of all schemes under a feedback deletion model: in this case, we only show the scenario with $\rho=0.5$, as performance is almost identical to the feedback erasure case. The only noticeable difference is that \gls{delta}+ degrades even faster, while the difference with the erasure model is negligible for all other schemes.

\section{Conclusion and Future Work}\label{sec:conc}

In this work, we presented \gls{delta}, a protocol that allows distributed sensor nodes to report anomalies efficiently by relying on the \gls{del} principle of common knowledge information. The protocol considerably outperforms both random access and scheduled schemes under reasonable operating conditions, and its operation is robust to relatively large shifts in its most significant parameter settings, as well as to imperfect feedback and traffic load estimation errors. Furthermore, the performance gap widens as the number of nodes increases, making the protocol suitable for large sensor networks.

Our work also opens several possible extensions and research directions, from a case in which anomalies are modeled as a more complex $N$-state Markov process to a more complex case in which nodes have structured beliefs about their own and others' observations.

\bibliographystyle{IEEEtran}
\bibliography{IEEEabrv,biblio.bib}

\end{document}